\def\year{2021}\relax
\documentclass[letterpaper]{article}
\usepackage{aaai21}
\usepackage{times}
\usepackage{helvet}
\usepackage{courier}
\usepackage[hyphens]{url}
\usepackage{graphicx}
\usepackage{graphicx}
\usepackage{natbib}
\usepackage{caption}
\nocopyright
\usepackage{algorithm,algorithmic}
\usepackage[table,x11names]{xcolor}
\usepackage{amsmath,amsfonts,amssymb,amsbsy}
\usepackage{subfig}
\usepackage{multirow}
\usepackage{enumitem}
\usepackage{tabularx}

\newcolumntype{L}[1]{>{\raggedright\arraybackslash}p{#1}}
\newcolumntype{C}[1]{>{\centering\arraybackslash}p{#1}}
\newcolumntype{R}[1]{>{\raggedleft\arraybackslash}p{#1}}

\usepackage[pagebackref=true,breaklinks=true,letterpaper=true,colorlinks,bookmarks=false]{hyperref}

\title{Back--Projection Pipeline}
\author{
    Pablo Navarrete Michelini, \textsuperscript{\rm 1}
    Hanwen Liu, \textsuperscript{\rm 1}
    Yunhua Lu, \textsuperscript{\rm 1}
    Xingqun Jiang, \textsuperscript{\rm 1}
    \\
}
\affiliations{
    \textsuperscript{\rm 1}BOE Technology Co., Ltd.
}
\begin{document}

\maketitle

\begin{abstract}
  We propose a simple extension of residual networks that works simultaneously in multiple resolutions. Our network design is inspired by the iterative back--projection algorithm but seeks the more difficult task of learning how to enhance images. Compared to similar approaches, we propose a novel solution to make back--projections run in multiple resolutions by using a data pipeline workflow. Features are updated at multiple scales in each layer of the network. The update dynamic through these layers includes interactions between different resolutions in a way that is causal in scale, and it is represented by a system of ODEs, as opposed to a single ODE in the case of ResNets. The system can be used as a generic multi--resolution approach to enhance images. We test it on several challenging tasks with special focus on super--resolution and raindrop removal. Our results are competitive with state--of--the--arts and show a strong ability of our system to learn both global and local image features.
\end{abstract}

\section{Introduction}

\begin{figure}
  \centering
  \includegraphics[width=\linewidth]{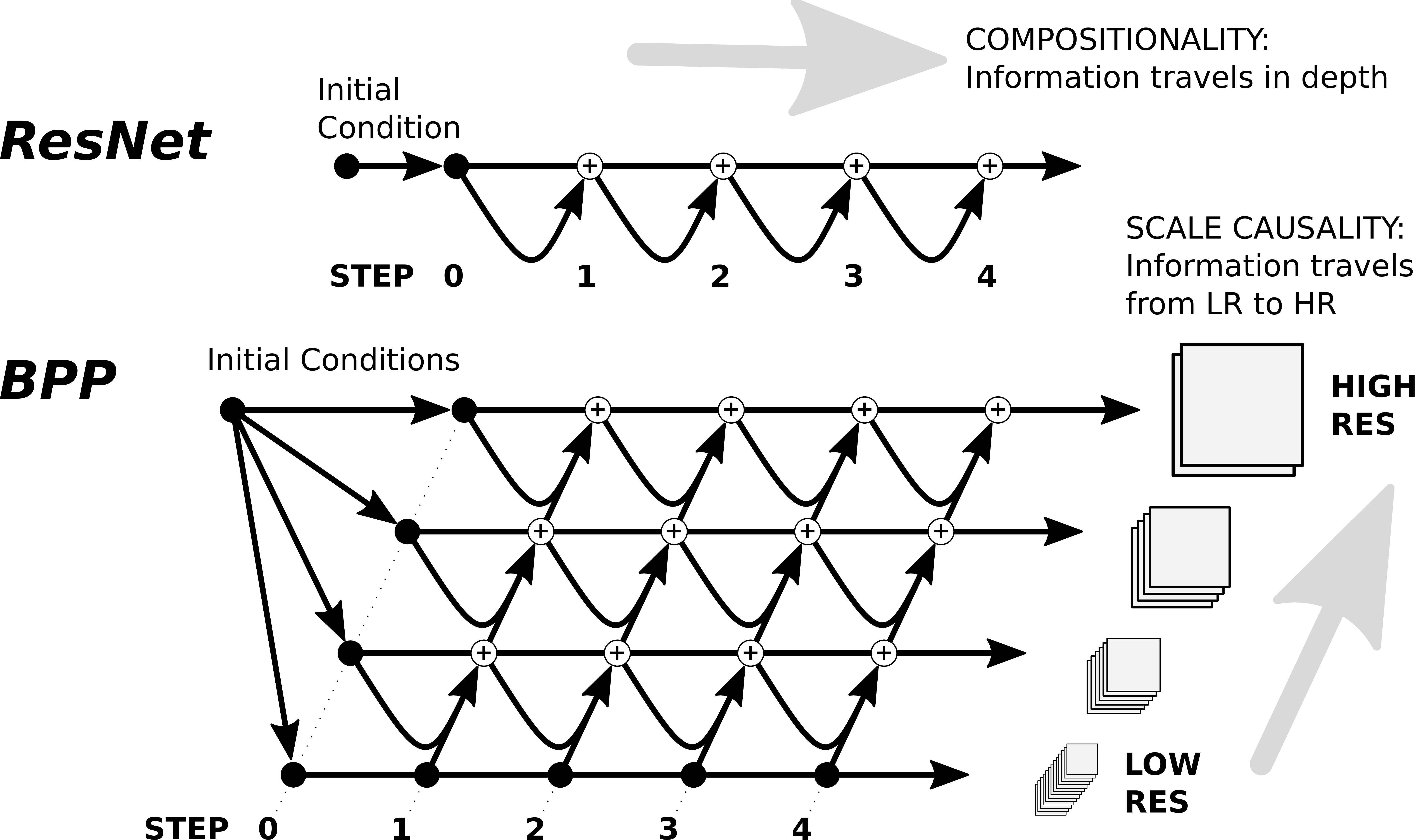}
  \caption{Our system (BPP) works as a multi--scale ResNet with state updates that interact with lower resolution states. Information travels forward in depth and upwards in scale.\label{fig:bpp_vs_resnet}}
\end{figure}

Image enhancement is the process of taking an impaired image as input and return an image of better quality. The current trend to achieve this target is to learn a mapping between impaired and enhanced images using example data. Deep--learning is leading this fast--growing quest in a number of applications, including: denoise\cite{lefkimmiatis2018universal}, deblur\cite{tao2018srndeblur}, super--resolution\cite{NTIRE2018_SR_Report}, demosaicking\cite{kokkinos2018deep}, compression removal\cite{lu2018deep}, dehaze\cite{ancuti2018ntire}, derain\cite{wang2019spatial}, raindrop removal\cite{Qian_2018_CVPR}, HDR\cite{Wu_2018_ECCV}, and colorization\cite{he2018deep}. Progress in network architectures often succeeds in image enhancement, as seen for example in image super--resolution, with CNNs applied in SRCNN \cite{Dong_2014a}, ResNets \cite{he2016deep} applied in EDSR \cite{Lim_2017_CVPR_Workshops}, DenseNets \cite{huang2017densely} applied in RDN \cite{zhang2018residual}, attention \cite{hu2018squeeze-and-excitation} applied in RCAN \cite{zhang2018rcan}, and non--local attention \cite{wang2018non-local} applied in RNAN \cite{zhang2019residual}. In all these examples, arguably the most influential practice is the use of residual networks (ResNets). Here, we define the \emph{network state} as the internal representation of an image in a network, commonly referred to as latent or feature space in the literature. The idea of ResNets is to represent an impaired image as a network state and progressively change it by adding residuals, as seen in Figure \ref{fig:bpp_vs_resnet}. This gives a compositional hierarchy\cite{poggio2017} of progressive local processing steps (e.g. convolutional layers) that transforms the input image. The update strategy of residual networks can be seen as a dynamical system where depth represents time and a differential equation models the evolution of the state\cite{liao2016bridging}.

Our proposed system, a \textbf{Back--Projection Pipeline (BPP)}, works as a residual network that carries many (instead of one) resolution states at a given time step as seen in Figure \ref{fig:bpp_vs_resnet}. Although similar in spirit to U--Nets \cite{ronneberger2015u-net}, this multi--resolution state is fundamentally different. U--Nets hold high resolution states to re--enter the network in later stages, whereas in BPP the state is created as initial conditions in multiple resolutions and get updated synchronously through the network. Another distinctive property of BPPs is \emph{scale causality}. Namely, after initialization, low resolution states do not depend on higher resolution states. Information travels forward in depth, same as in ResNets, and upwards in scale, as shown in Figure \ref{fig:bpp_vs_resnet}. Scale causality is inspired by scale--space \cite{lindeberg94} and multi--resolution analysis\cite{SMallat_1998a} to express the nested nature of details. A simple example is that when we see an image of a keyboard we expect to see letters, but not necessarily the other way around. Finally, the interpretation of BPPs as an extension of ResNets becomes more clear from the dynamic of the network. We will show that BPP updates can be modeled by a non--autonomous system of differential equations, as opposed to a single ODE for ResNets.

\textbf{Related Work}. With regard to applications, BPP gives us a generic multi--resolution approach to transform images into a desired target. Current benchmarks in image enhancement often use different architectures for different tasks. It is important to distinguish between local and global targets. In the problem of super--resolution, for example, we need to calculate pixel values around a local area, and distant pixels become less relevant. In a different problem, contrast enhancement, we want to change the histogram of an image, which contains statistics that represent global features. General image enhancement is gaining interest in research and has been considered in the context of:
\begin{itemize}[leftmargin=*]
    \item \emph{Mixed Local Problems}: In \cite{zhang2019residual}, for example, authors solve denoising, super--resolution and deblur tasks using a single architecture and different parameters for each problem. In \cite{Gharbi2016,ehret2019joint} authors solve joint demosaicking and denoising, and in \cite{Qian2019} authors additionally solve super-resolution, all through using a single architecture and same model parameters. In \cite{zhang2018learning} authors tackle super--resolution and deblur, and train a single system to handle different image degradations.
    \item \emph{Global and Local Problems}: Authors in \cite{soh_joint_2019,Kim2019,kinoshita2019} consider the joint solution of low--to--high dynamic range enhancement as well as image--SR. In \cite{Kim2019} authors generate an image in HDR display format, whereas \cite{soh_joint_2019,kinoshita2019} use the same input and output format. They both use U--Net configurations, while \cite{soh_joint_2019} uses a two--stage Retinex decomposition network.
\end{itemize}

Regarding architecture, BPP uses a multi--resolution workflow, which is different from U--Nets \cite{ronneberger2015u-net}. This workflow follows from the Iterative Back--Projection \cite{Irani_1991a} (IBP) algorithm. In this respect, Multi--Grid Back--Projection \cite{PNavarrete_2019a} (MGBP) is the closest super--resolution system that is state--of--the--arts for lightweight systems with small number of parameters. It is based on a multi--resolution back--projection algorithm that uses a multigrid recursion\cite{UTrottenberg_2000a}. This recursion violates scale--causality as it sends network states back to low--resolution to restart iterations. We also notice that BPP follows the wide--activation design in \cite{yu2018wide} in the sense that features are increase before activations and reduced before updating. BPP shows a workflow structure similar to the Multi--scale DenseNet architecture in \cite{huang2018multi-scale}, except in the latter scale--causality moves downwards in scale, it does not use back--projections and it focuses on a label prediction problem. The WaveNet architecture \cite{oord2016wavenet} also shares the property of scale causality but without back--projections, moving information upwards in scale without any step back. Finally, a similar causality and adaptation in the number of channels per scale exists in the SlowFast architecture \cite{Feichtenhofer_2019_ICCV} but again without back--projections.

\textbf{Contributions}. Our major contribution is the introduction of a new network architecture that extends ResNets from single to multiple resolutions, with a clear representation in terms of ODE dynamic. Our main focus is to evaluate this extension and to prove that it is beneficial with respect to conventional ResNets. We also verify that the multi--scale dynamic of the network is being used to achieve improved performance and we visualize the dynamic of the network in solving different problems. BPP can be used to solve joint local problems, as well as combinations of global and local problems, getting state--of--art results in image--SR and competitive results for other problems using a single network configuration. Finally, we also show empirical evidence that BPP effectively uses both local and global information to solve problems.
\begin{figure}
  \centering
  \includegraphics[width=\linewidth]{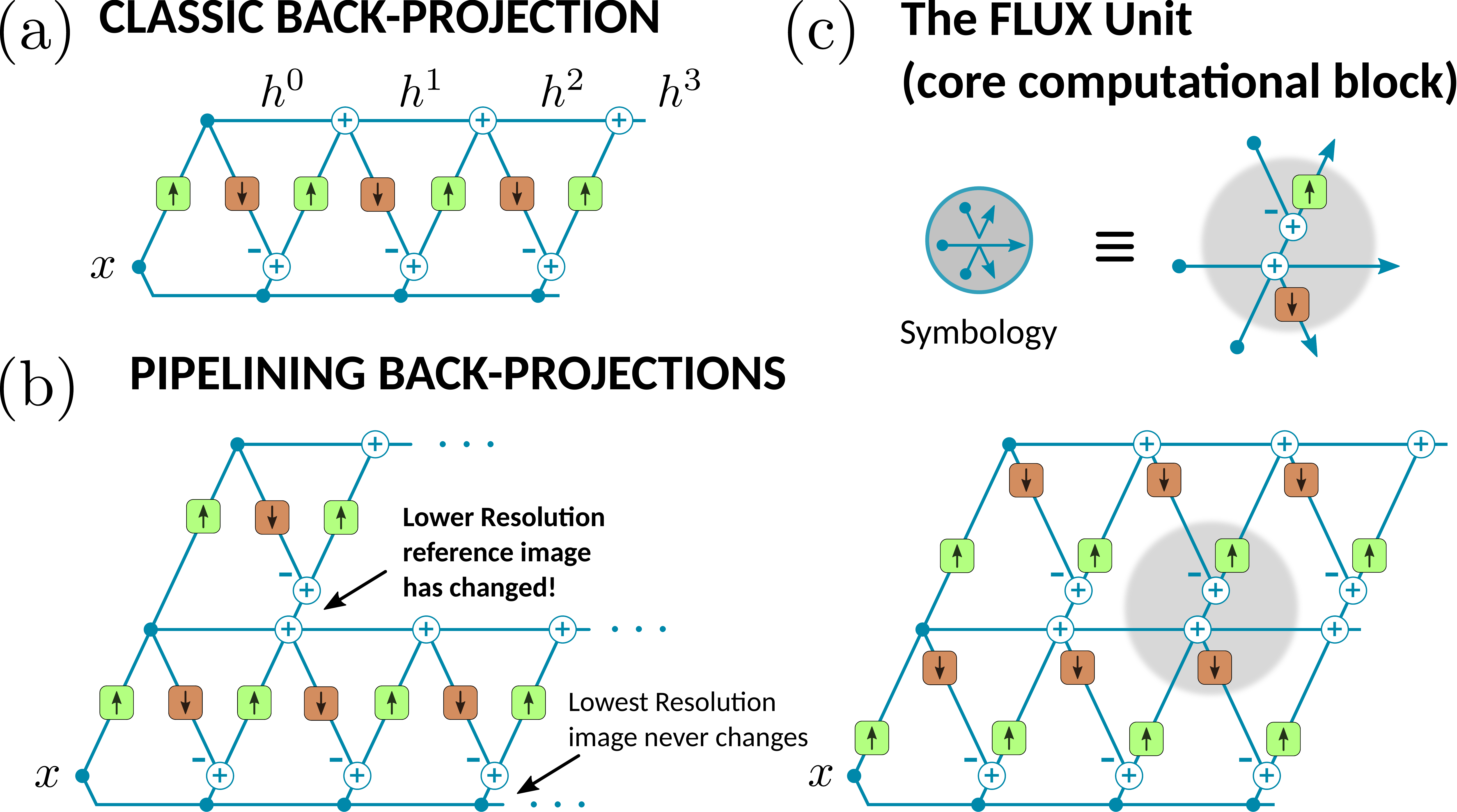}
  \caption{Pipelining Iterative Back--Projections. \label{fig:pipelining}}
\end{figure}

\begin{figure*}[t]
  \centering
  \includegraphics[width=\linewidth]{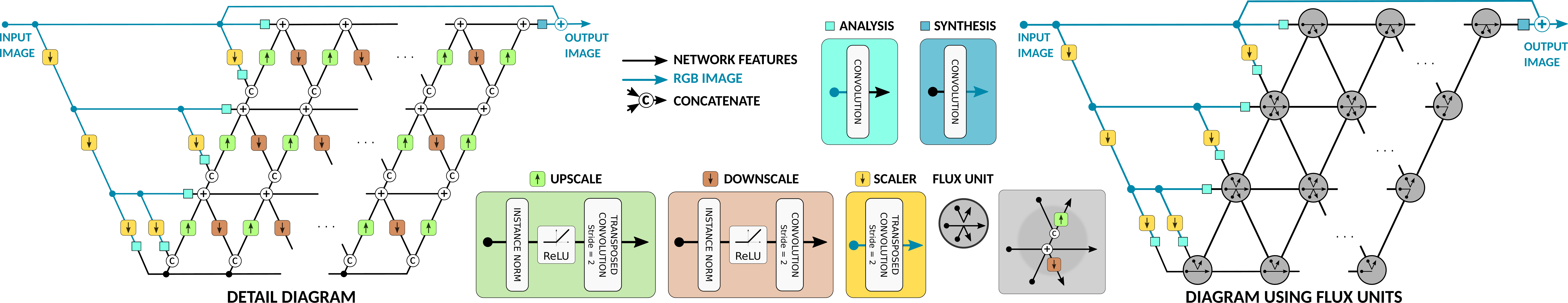}
  \caption{Back--Projection Pipeline network diagram. On the left, a detailed diagram shows all back--projection modules. On the right, the diagram is simplified by using \emph{Flux} units. \label{fig:bpp}}
\end{figure*}
\begin{algorithm*}[t]
    \centering
    \begin{tabular}{lll}
        \emph{BPP}$(input,L,D)$: & \emph{FluxBlock}$(x_k,p_k,L)$: & \emph{Flux}$(e_{in}, x_{in}, p_{in})$: \\

        \resizebox{.3333\textwidth}{!}{
            \begin{minipage}{.43\textwidth}
            \begin{algorithmic}[1]
                \REQUIRE Image $input$.
                \REQUIRE Integer $L\geqslant 1, D\geqslant 1$.
                \ENSURE Image $output$.

                \STATE $s^A_L = input$
                \FOR{$k = L-1,\ldots,1$}
                    \STATE $s^A_k = Scaler^A_k(s^A_{k+1})$
                \ENDFOR
                \STATE $x_L = Analysis^A_k(s^A_L)$
                \FOR{$k = 1,\ldots,L-1$}
                    \STATE $x_k = Analysis^A_k(s^A_k)$
                    \STATE $s^B_k = Scaler^B_k(s^A_{k+1})$
                    \STATE $p_k = Analysis^B_k(s^B_k)$
                \ENDFOR

                \FOR{$l = 1,\ldots,D$}
                    \STATE $x, p = FluxBlock(x, p, L)$
                \ENDFOR
                \STATE $output = input + Synthesis(x_L)$
            \end{algorithmic}
        \end{minipage}
        }

        &

        \resizebox{.3333\textwidth}{!}{
            \begin{minipage}{.43\textwidth}
            \vspace{0.05in}
            \begin{algorithmic}[1]
                \REQUIRE Initial $x_k, p_k$, $k = 1,\ldots,L$.
                \REQUIRE Integer $L\geqslant 1$.
                \ENSURE Updated $x_k, p_k$, $k = 1,\ldots,L$.

                \STATE $e_{2}, x_1, \_ = Flux(0, x_1, p_1)$
                \FOR{$k = 2,\ldots,L$}
                    \STATE $e_{k+1}, x_k, p_{k-1} = Flux(e_k, x_k, p_k)$
                \ENDFOR
                \STATE $\_, x_L, p_{L-1} = Flux(e_L, x_L, 0)$
            \end{algorithmic}
            {\vspace{.01in}\hspace{.1\linewidth}\includegraphics[width=.8\linewidth]{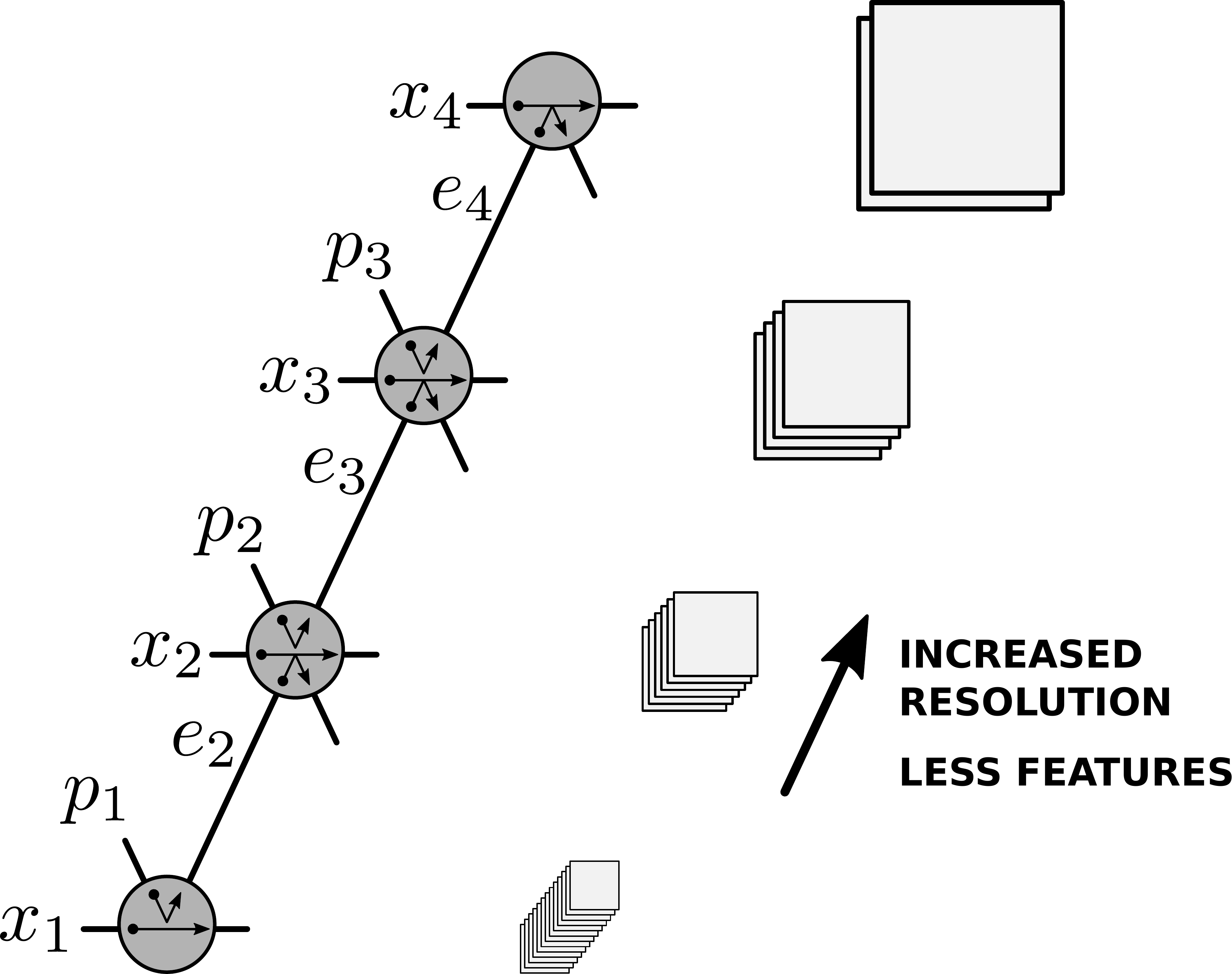}}
        \end{minipage}
        }

        &

        \resizebox{.3333\textwidth}{!}{
            \begin{minipage}{0.43\textwidth}
            \vspace{-0.3in}
            \begin{algorithmic}[1]
                \REQUIRE $e_{in}, x_{in}, p_{in}$.
                \ENSURE $e_{out}, x_{out}, p_{out}$.

                \STATE $c = x_{in} + e_{in}$
                \STATE $e_{out} = Upscale([\;p_{in}, c\;])$
                \STATE $p_{out} = Downscale(c)$
                \STATE $x_{out} = Update(c)$
            \end{algorithmic}
            {\vspace{.3in}\hspace{.1\linewidth}\includegraphics[width=.5\linewidth]{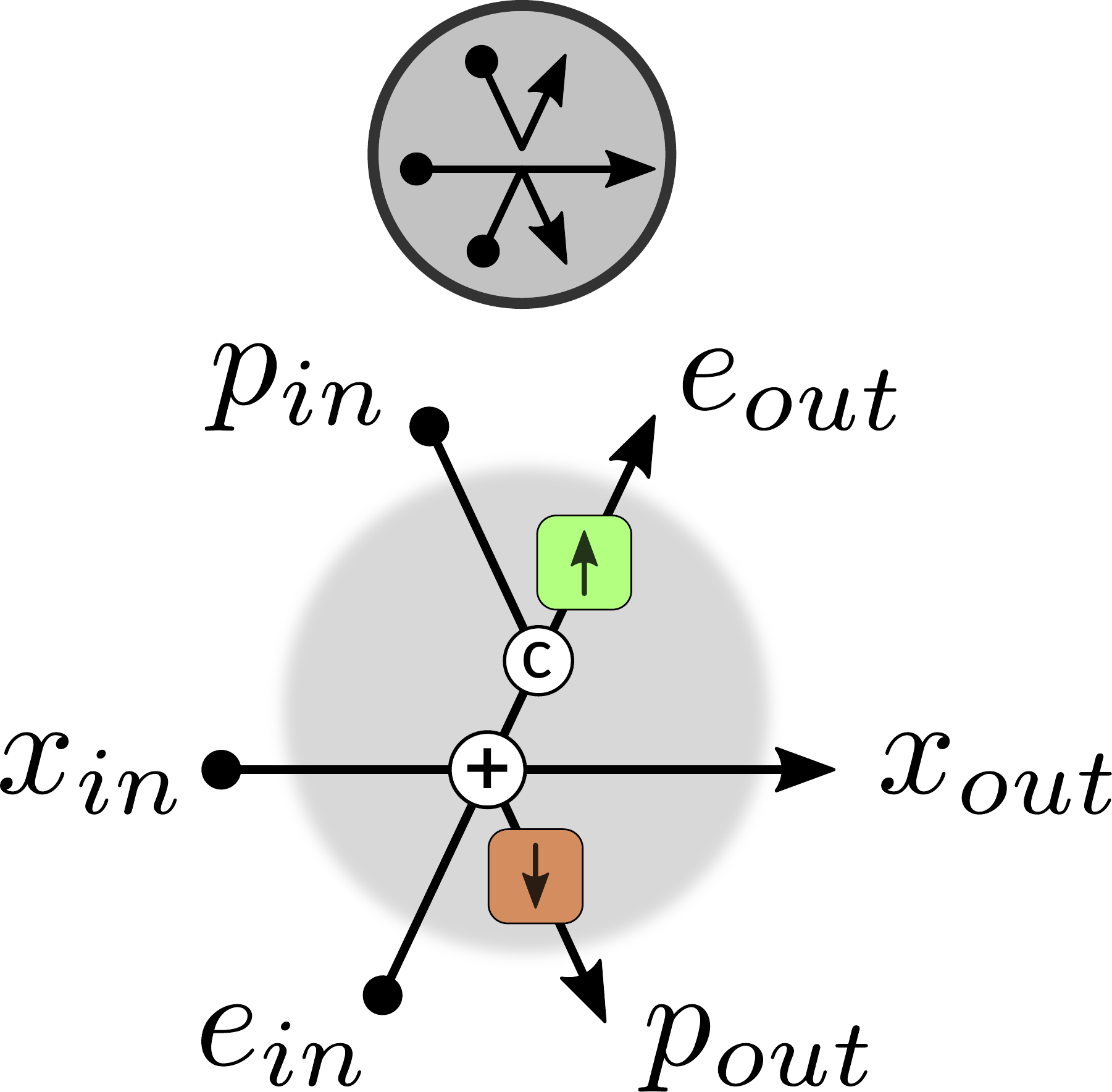}}
            \end{minipage}
        }
    \end{tabular}
    \caption{Back--Projection Pipeline (BPP)} \label{tab:alg}
    \label{alg:bpp}
\end{algorithm*}

\section{Architecture Design}
In Figure \ref{fig:pipelining} (a) we observe the workflow of the Iterative Back--Projections \cite{Irani_1991a} (IBP) algorithm:
\begin{align}
    h^0 &= P \; x\;, & h^{t+1} &= h^t + P\; e(h^t) \;,\nonumber \\
        &                & e(h^t)  &= x - R\; h^t \;.
\end{align}
IBP upscales an image $x$ with a linear operator $P$ and sends it back to low--resolution to verify the downscaling model represented by a linear operator $R$. Now, we propose to extend the IBP algorithm to multiple scales by using the data pipeline approach shown in Figure \ref{fig:pipelining} (b). Specifically, as soon as we get the first upscale image, we take it as reference and start a new upscaling to a higher resolution. Next, we downscale the second high--resolution image to verify the downscaling model. However, the reference image has been changed by the back--projection update at the lower level. At the lowest resolution the image never changes, and upper level iterations need to keep track of the lower level updates. In Figure \ref{fig:pipelining} (b) we identify the essential computational block to assemble the pipeline: the \emph{Flux} unit. The \emph{Flux} unit is what makes scale travel possible by connecting input and output images from different levels.

\textbf{Network Architecture}. Without loss of generality, we tackle the image enhancement problem with an input resolution equal to the output resolution. In the case of image SR, which requires to increase image resolution, we add a pre--processing stage where the input image is upscaled using a standard method (e.g. bicubic). This helps to make the system become more general for applications. For example, we can easily solve the problem of fractional upscaling factors\cite{hu2019meta-sr} or multiple upscaling factors\cite{zhang2018learning} by simply using different pre--processing bicubic upscalers.

The full BPP algorithm and network configuration is specified in Algorithm \ref{alg:bpp} and Figure \ref{fig:bpp}. To extend the pipelining approach into a network configuration, first, we initialize the network states $x_k$ and down--projections $p_k$ using linear downscalers and single convolutional layers in the \emph{Analysis} modules to increase the number of channels. Second, states are updated using the \emph{Flux--Blocks} defined in Algorithm \ref{alg:bpp}, calculating residuals $e_k$ and updating states upwards in scale with flux units. Third, the output state in the highest resolution is converted into a residual image by a convolutional layer in the \emph{Synthesis} module and added to the input image.

\begin{figure}
  \centering
  \includegraphics[width=.4\linewidth]{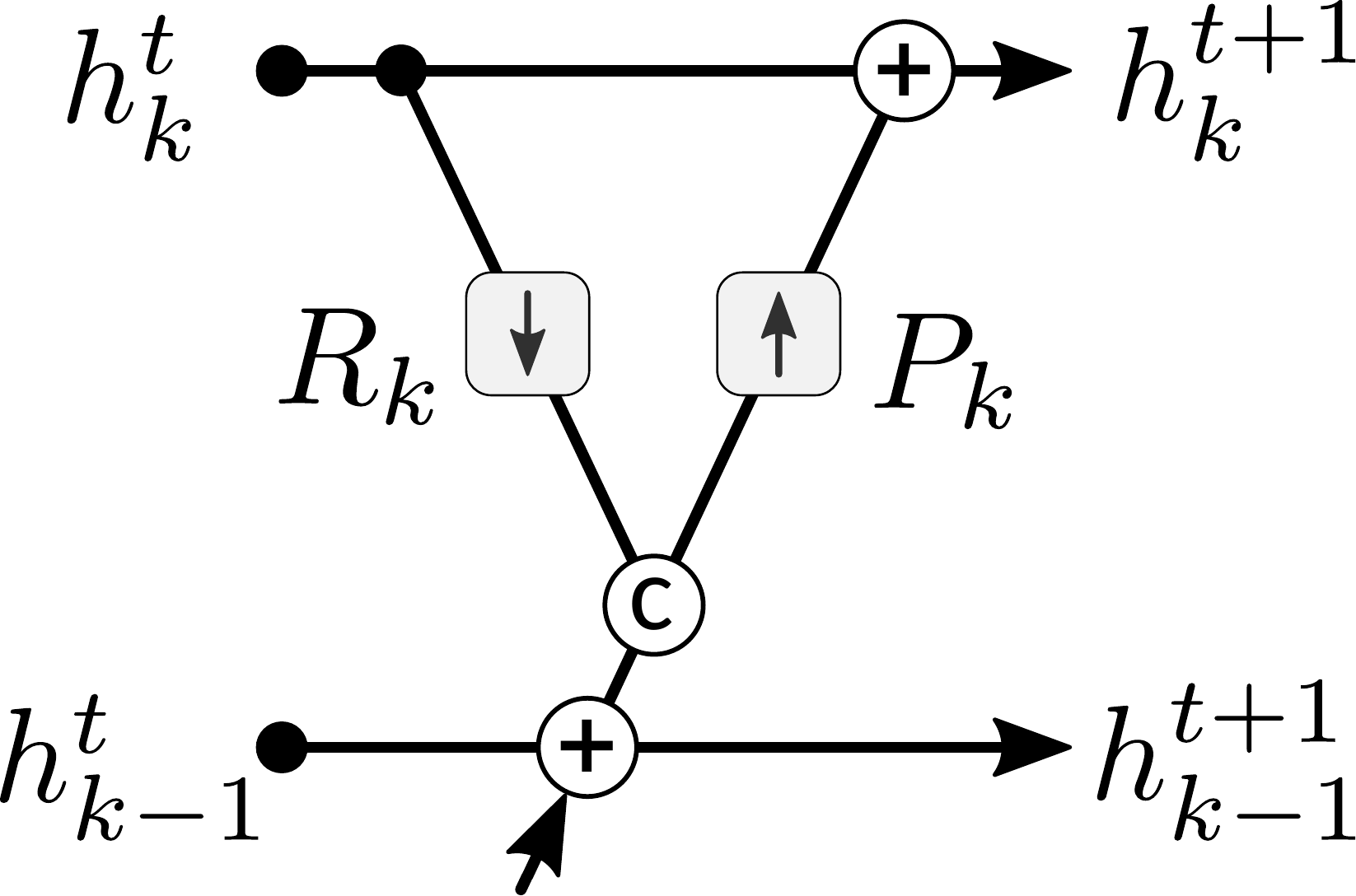}
  \caption{
State diagram of the depth transitions in the BPP architecture. The residual structure leads to a non--autonomous system of differential equations. \label{fig:dynamic}}
\end{figure}

\textbf{Network Dynamic}. The restriction operators $R_k$ (\emph{Downscale} module) and interpolation operators $P_k$ (\emph{Upscale} module) are now non--linear and do not share parameters (time dependent). When we interpret depth as time $t$, the dynamic is described in Figure \ref{fig:dynamic} and leads to the following set of difference equations with their correspondent extension to continuous time:
\begin{align}
& & h_k^{t+1}       &= h_k^t + P_k(R_k(h_k^t, t), h_{k-1}^{t+1}, t) \nonumber \\
& & h_1^{t+1}       &= h_1^t \;, \nonumber \\
& \stackrel{cont.time}{\Rightarrow} & \nonumber \\
& & \frac{dh_k}{dt} &= P_k(R_k(h_k, t), h_{k-1}, t) \nonumber \\
& & h_1(x, y, t)    &= h_1(x, y, 0) \;. \label{eq:bpp_dynamic}
\end{align}
In the case of ResNets, the dynamical systems is given by $h^{t+1} = h^t + f(h^t, t)$ and $\tfrac{dh}{dt}=f(h,t)$ in continuous time. Therefore, BPP extends the model of ResNets from a single ODE to a system of coupled equations. Scale--causality follows from \eqref{eq:bpp_dynamic} as state $h_k$ only depends on $h_{k-1}, h_{k-2},\ldots$. The multi--scale nature follows from the spatial dimension of state vectors $h_k$, explicitly expressed in operators $P_k:\mathbb{R}^{\frac{H}{2}\times\frac{W}{2}}\rightarrow\mathbb{R}^{H\times W}$ and $R_k:\mathbb{R}^{H\times W}\rightarrow\mathbb{R}^{\frac{H}{2}\times\frac{W}{2}}$. In continuous space we could also express the multi--scale nature of the equations by using initial conditions $h_{k-s}(x, y, t=0) = h_k(2^s x, 2^s y, t=0)$ with $s\in\mathbb{N}$ with no filtering needed in continuous space, since aliasing effects do not exist. We observe that initial conditions are self-similar in scale\cite{SMallat_1998a}. Whether this property is maintained in time depends on the evolution of the network state. In the continuous time model, the restriction operator $R_k$ in \eqref{eq:bpp_dynamic} represents a renormalization--group transformation of the network state, similar to those used in particle physics and ODEs to ensure self--similarity\cite{fisher1974the,chen1996renormalization}. In this sense, using different parameters at each scale allows the model to adjust the level of self--similarity that works better for a given problem. On the other hand, using different parameters in time can also be beneficial. It has been observed in \cite{liao2016bridging} that normalization layers do not work well in recurrent networks, which share parameters in time. But in time--dependent systems, these layers become beneficial. Since the BPP configurations in our experiments use IN--layers, we chose to use different parameters in time. This does not have a significant effect in performance, because the flux--block structure in Algorithm \ref{alg:bpp} uses inline updates that avoid storage of old network states. During training, a checkpoint strategy can effectively reduce the memory footprint \cite{chen2016training}.

Using pipelining to extend IBP into multiple scales is simple and this is the major strength of this approach. There are several ways to extend IBP to multiple scales. We mentioned MGBP as a relevant but different approach. BPP is simpler, and that simplicity translates to a clear ODE model that is difficult to obtain otherwise. Most importantly, this ODE model is very expressive about the connection to IBP. It is direct from \eqref{eq:bpp_dynamic} that if the composition of $P$ and $R$ operations forms a contraction mapping then the ODE model will converge, which is the same argument used in convergence proofs of IBP in the linear case \cite{Irani_1991a}. At this point BPP departs from IBP. Because BPP is trained in a supervised fashion, we do not know a priori how is this dynamic going to be driven towards the target. Overall, the BPP model inherits the essence of IBP in terms of an iteration that updates residuals upwards in scale, which can now be trained to reach diverse targets in a non--linear fashion using convolutional networks. The main purpose of our investigation is: first, to generalize the IBP dynamic to multiple scales in sequence; and second, to study how powerful is this dynamic so solve more general problems.

Finally, we note that the continuous model in \eqref{eq:bpp_dynamic} allows BPP to work as a Neural--ODE system\cite{chen2018neural}. For the sake of simplicity, in this work we do not explore this direction.
However, it stands as an interesting direction for future research.

\begin{figure*}[t!]
  \centering
  \includegraphics[width=\linewidth]{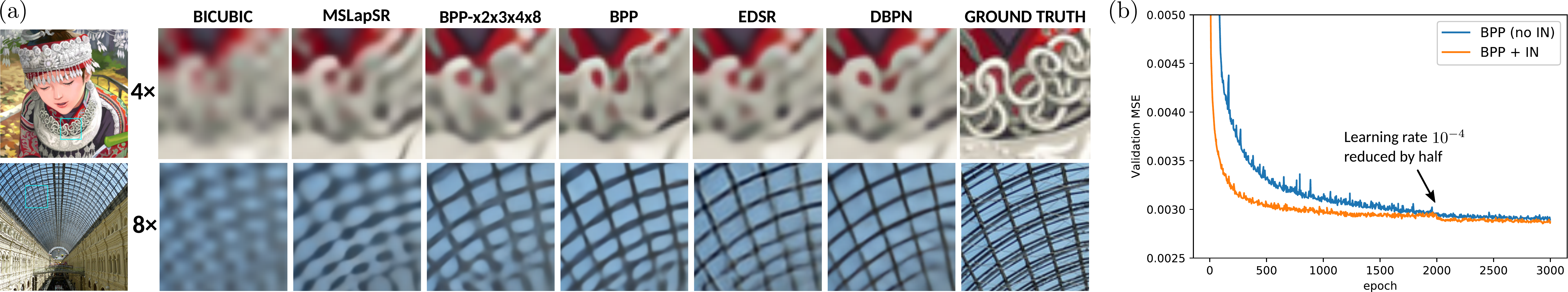}
  \caption{a) Qualitative evaluation for SR methods. b) Validation MSE for $4\times$ SR. \label{fig:sr}}
\end{figure*}
\begin{table*}
\caption{Quantitative evaluation for SR. A more extensive comparison is available in the Appendix.} \label{tab:sr}
\centering
\resizebox{\linewidth}{!}{
\begin{tabular}{lC{3cm}C{2cm}C{2cm}C{2cm}C{2cm}C{2cm}C{2cm}C{2cm}C{2cm}}
    \hline
    & &
    \multicolumn{2}{c}{Set14} &
    \multicolumn{2}{c}{BSDS100} &
    \multicolumn{2}{c}{Urban100} &
    \multicolumn{2}{c}{Manga109} \\
    Algorithm & & PSNR--$Y_M$ & SSIM--$Y_M$ & PSNR--$Y_M$ & SSIM--$Y_M$ & PSNR--$Y_M$ & SSIM--$Y_M$ & PSNR--$Y_M$ & SSIM--$Y_M$ \\
    \hline
    Bicubic & \multirow{6}{*}{$2\times$}                           &  30.34  &  0.870  &  29.56  &  0.844  &  26.88  &  0.841  &  30.84  &  0.935  \\
    MSLapSRN                                                     & &  33.28  &  0.915  &  32.05  &  0.898  &  31.15  &  0.919  &  37.78  &  0.976  \\
    D-DBPN                                                       & &  33.85  &  0.919  &  32.27  &  0.900  &  32.70  &  0.931  &  39.10  &  0.978  \\
    EDSR                                                         & &  33.92  &  0.919  &  32.32  &  0.901  &  32.93  &  0.935  &  39.10  &  0.977  \\
    RDN                                                          & & \textbf{34.28} & \textbf{0.924} & \textbf{32.46} & \textbf{0.903} & \textbf{33.36} & \textbf{0.939} & \textbf{39.74} & \textbf{0.979} \\
    \rowcolor{lightgray} BPP--SRx2x3x4x8                         & &  33.27  &  0.913  &  31.21  &  0.879  &  31.67  &  0.921  &  38.31  &  0.975  \\
    \rowcolor{lightgray} BPP--SRx2                               & &  34.23  &  0.922  &  31.63  &  0.886  &  33.07  &  0.935  &  39.19  &  0.977  \\
    \noalign{\smallskip}\hline\noalign{\smallskip}
    Bicubic & \multirow{5}{*}{$3\times$}                           &  27.55  &  0.774  &  27.21  &  0.739  &  24.46  &  0.735  &  26.95  &  0.856  \\
    MSLapSRN                                                     & &  29.97  &  0.836  &  28.93  &  0.800  &  27.47  &  0.837  &  32.68  &  0.939  \\
    EDSR                                                         & &  30.52  &  0.846  &  29.25  &  0.809  &  28.80  &  0.865  &  34.17  &  0.948  \\
    RDN                                                          & &  30.74 & \textbf{0.850} & \textbf{29.38} & \textbf{0.812} & 29.18 & 0.872 & \textbf{34.81} & \textbf{0.951} \\
    \rowcolor{lightgray} BPP--SRx2x3x4x8                         & &  30.23  &  0.838  &  28.81  &  0.794  &  28.43  &  0.852  &  33.75  &  0.943  \\
    \rowcolor{lightgray} BPP--SRx3                               & &  \textbf{30.78}  &  0.848  &  29.14  &  0.804  &  \textbf{29.56}  &  \textbf{0.873}  &  34.49  &  0.948  \\
    \noalign{\smallskip}\hline\noalign{\smallskip}
    Bicubic & \multirow{6}{*}{$4\times$}                           &  26.10  &  0.704  &  25.96  &  0.669  &  23.15  &  0.659  &  24.92  &  0.789  \\
    MSLapSRN                                                     & &  28.26  &  0.774  &  27.43  &  0.731  &  25.51  &  0.768  &  29.54  &  0.897  \\
    D-DBPN                                                       & &  28.82  &  0.786  &  27.72  &  0.740  &  26.54  &  0.795  &  31.18  &  0.914  \\
    EDSR                                                         & &  28.80  &  0.788  &  27.71  &  0.742  &  26.64  &  0.803  &  31.02  &  0.915  \\
    RDN                                                          & &  29.01  &  \textbf{0.791}  &  27.85  & \textbf{0.745} &  27.01  &  0.812  & \textbf{31.74} & \textbf{0.921} \\
    \rowcolor{lightgray} BPP--SRx2x3x4x8                         & &  28.55  &  0.778  &  27.43  &  0.728  &  26.48  &  0.791  &  30.81  &  0.909  \\
    \rowcolor{lightgray} BPP--SRx4                               & &  \textbf{29.07}  &  \textbf{0.791}  &  \textbf{27.89}  &  \textbf{0.745}  &  \textbf{27.55}  &  \textbf{0.819}  &  31.63  &  0.918  \\
    \noalign{\smallskip}\hline\noalign{\smallskip}
    Bicubic & \multirow{6}{*}{$8\times$}                           &  23.19  &  0.568  &  23.67  &  0.547  &  20.74  &  0.516  &  21.47  &  0.647  \\
    MSLapSRN                                                     & &  24.57  &  0.629  &  24.65  &  0.592  &  22.06  &  0.598  &  23.90  &  0.759  \\
    D-DBPN                                                       & &  25.13  &  0.648  &  24.88  &  0.601  &  22.83  &  0.622  &  25.30  &  0.799  \\
    EDSR                                                         & &  24.94  &  0.640  &  24.80  &  0.596  &  22.47  &  0.620  &  24.58  &  0.778  \\
    RDN                                                          & &  25.38  &  0.654  &  25.01  &  0.606  &  23.04  &  0.644  &  \textbf{25.48}  &  \textbf{0.806}  \\
    \rowcolor{lightgray} BPP--SRx2x3x4x8                         & &  25.10  &  0.642  &  24.89  &  0.598  &  22.72  &  0.626  &  24.78  &  0.785  \\
    \rowcolor{lightgray} BPP--SRx8                               & &  \textbf{25.53}  &  \textbf{0.655}  &  \textbf{25.11}  &  \textbf{0.607}  &  \textbf{23.17}  &  \textbf{0.649}  &  25.28  &  0.800  \\
    \hline
    \end{tabular}
}
\end{table*}

\section{Experiments}
In our experiments we found that using IN--layers to activate ReLU units, as shown in Figure \ref{fig:bpp}, could help converge faster in early training and doing so independent of initialization. Figure \ref{fig:sr} (b) shows this effect and we also see that IN--layers are not required for BPP in the long run. In early training IN layers placed before ReLUs force a $50\%$ activation in all flux units across all scales. This strategy shows to be a good choice to initialize parameters. Alternatively, we found that the most effective way to avoid IN--layers is using Dirac--kernels to initialize weights and adding Gaussian noise. This initialization was used in the learning curve \emph{BPP (no IN}) in Figure \ref{fig:sr} (b) and it is the closest we have found to avoid normalization layers.

Because of memory limitations we used a patch--based training strategy, where smaller--sized patches are taken from training set images. Patch--based learning reduces the receptive field of the network during training. At inference the performance of the network reduces if the mean and variance of IN--layers are computed on an image larger than the training patches. To solve this problems we: first, divide input images into overlapping patches (of same size as training patches); second, we multiply each output by a Hamming window\cite{harris1978use}; and third, we average the results. In all our experiments we use overlapping patches separated by $16$ pixels in vertical and horizontal directions. The weighted average helps to avoid blocking artifacts.

On one hand, this approach introduces redundancy and reduces performance for medium size images. On the other hand, it also allows the algorithm to run on very large images (e.g. 8K) and can be massively parallelized by batch processing in multiple GPUs.

\textbf{Configuration}. In the following experiments we use a BPP configuration with $16$ back--projection layers (flux--blocks), $4$ resolution levels and $256$, $128$, $64$ and $48$ features per level from lowest to highest resolution, respectively. All convolutional layers use $3\times 3$ as kernel size, and scalers are initialized with bicubic filters of size $9\times 9$ and trained as additional parameters. A fully unrolled diagram is shown in the Appendix. The configuration was tuned according to validation performance for the most challanging problems (e.g. SR--$8\times$). By fixing the configuration we can potentially have the architecture hardwired in silicon and update its model parameters to switch between different problems.

\textbf{Performance}. The BPP architecture is multi--scale and sequential. The so--called \emph{Flux--Block} in Algorithm \ref{alg:bpp} represents the sequential block and consists of one Flux unit per level. This sequential structure is more convenient for memory performance as it avoids buffering of features from previous blocks. Architectures such as Dense--Nets, U--Nets and MGBP need to buffer features in skipped connections and thus need more memory. Because the configuration is fixed, the performance of the system can be roughly estimated from average statistics. The system has a total of $19$ million parameters and it can process $1.7$ million pixels per second on a Titan X GPU using $16$--bit floating point precision. This means, for example, that it takes $3.7$ seconds to process a Full--HD image in RGB format ($3\times 1920\times 1080$ pixels).

\textbf{P1: Image Super--Resolution}. We use DIV2K \cite{Agustsson_2017_CVPR_Workshops} and FLICKR--2K datasets for training and the following datasets for test: Set--14\cite{zeyde2010on}, BSDS--100\cite{martin2001a}, Urban--100\cite{huang2015single} and Manga--109\cite{matsui2017sketch-based}. Impaired images were obtained by downscaling and then upscaling ground truth images, using Bicubic scaler, and scaling factors: $2\times$, $3\times$, $4\times$ and $8\times$. Here, we consider two cases: we trained models BPP--SR$\times f$ for each upscaling factor $f=2, 3, 4$ and $8$; and we also trained a single model BPP--SRx2x3x4x8 to restore impaired images with unknown upscaling factors. We use $16$ patches per mini--batch with patch size $48f\times 48f$ for known upscaling factor $f$, and $192\times 192$ for unknown upscaling factor, all at high resolution.

Table \ref{tab:sr} and Figure \ref{fig:sr} (a) show quantitative and qualitative results compared to other methods. We focus our comparison to the following methods: Bicubic (the baseline); EDSR \cite{Lim_2017_CVPR_Workshops}, with major processing in $1$ resolution level using a $32$--layer ResNet; Dense--DBPN \cite{DBPN2018}, with major processing in $2$ resolution levels using $12$ densely connected up/down back--projections; and RDN \cite{zhang2018rdnir}, with major processing in $1$ resolution level using $20$ densely connected residual--dense--blocks. We show EDSR and DBPN because they are both closely related to BPP in their residual and back--projection structures, respectively, and we show RDN as a top reference of current state--of--the--arts. Further comparisons with other methods can be found in the Appendix.

Overall, for the problem of super--resolution we find that BPP can get excellent results, reaching state--of--the--arts results in both quantitative and qualitative evaluations, but it decreases its performance when we test a more general problem. First, BPP--$\times f$ models get the best scores in most quantitative and qualitative evaluations, with RDN slightly outperforming BPP at $2\times$ and $3\times$ upscaling factors. This setting, including datasets for training and test, is the most common evaluation procedure for supervised SR technics. In terms of application this would be useful if we need to enhance an image upscaled with Bicubic upscaler with a specific upscaling factor. It often happens that we have an image upscaled with an unknown factor and in this case we do not know which model parameters to load. In this case the BPP--SRx2x3x4x8 model offers a general upscaling solution. This performance of these BPP models decrease and not reach state--of--the--arts results. Although reasonably close to state--of--the--arts, often outperforming EDSR, we would have expected this model to perform better than BPP--$\times f$ if the architecture was able to generalize effectively to this more general setting. In fact, it has been observed in VDSR \cite{Kim_2016_VDSR} and MDSR \cite{Lim_2017_CVPR_Workshops} that training with unkown upscaling factors can improve the performance of the network. Therefore, these empirical results show that BPP can be very effective for fixed upscaling factors but does not generalize as well as other architectures for general upscaling factors.

\textbf{P2: Raindrop Removal}. We use the DeRaindrop\cite{Qian_2018_CVPR,deraindrop-dataset} dataset for training and test. This dataset provides paired images, one degraded by raindrops and the other one free from raindrops. These were obtained by use two pieces of exactly the same glass: one sprayed with water, and the other is left clean. In each training batch, we take $1$ patch of size $528\times 528$. We train a BPP model using $L_1$ loss and patch size $456\times 456$. More details of training settings are provided in the Appendix.

This problem is very different in nature to super--resolution. On one hand, a significant portion of pixels contain (uncorrupted) high--resolution information that must move to the output with little or no change. At the same time it needs to identify the irregular distribution of raindrops, with different sizes, and fill--in those areas by predicting the content within. In some images the content within raindrops is of little use, making the problem similar to inpainting. Thus, the problem requires processing of both local and global information in order to fill--in raindrops.

\begin{figure}
  \centering
  \caption{Qualitative evaluation for raindrop removal. \label{fig:raindrop}}
  \includegraphics[width=\linewidth]{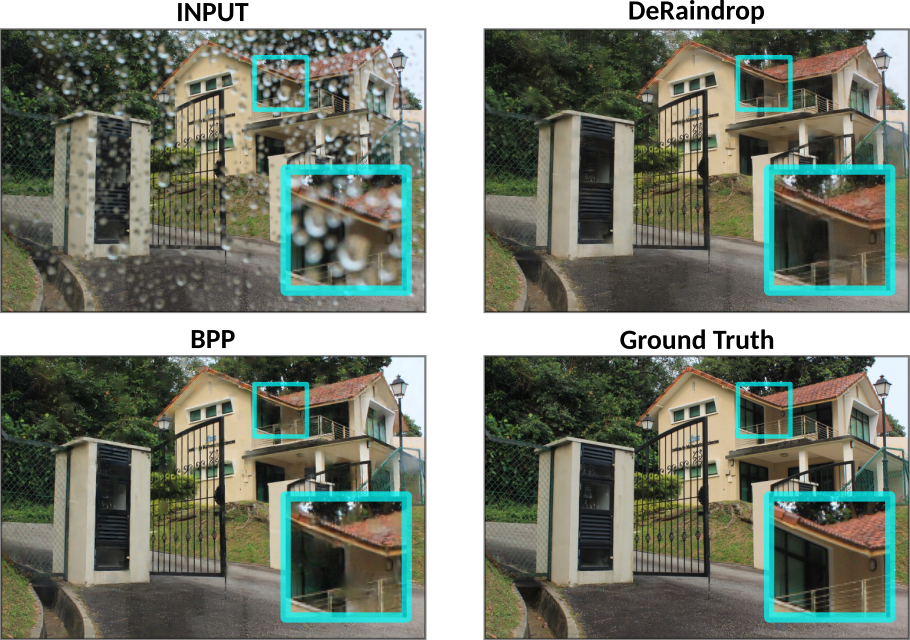}
\end{figure}

Even though we only trained our system with an $L_1$ loss, our system performs similar to the state--of--the--arts DeRaindrop \cite{Qian_2018_CVPR} as seen in Table \ref{tab:raindrop} and Figure \ref{fig:raindrop}. The DeRaindrop network in \cite{Qian_2018_CVPR} uses an attentive GAN approach that can estimate raindrop masks to focus on these areas for restoration. The PSNR score of BPP is better than DeRaindrop without adversarial training, and the SSIM score is better than all other systems in Table \ref{tab:raindrop}. The qualitative evaluation shows that BPP achieves a reasonable quality, considering the fact that it has not been trained using GANs. Here, the BPP architecture appears to be effective. In the next section we inspect properties of the network that reveal the undergoing mechanism used by BPP to obtain its solutions.

\textbf{Other Problems}. The performance in other problems, including mobile--to--DSLR photo translation, dehaze and joint HDR+SR are included in the Appendix.

\begin{table}
\normalsize
\centering
\vspace{-0.1in}
\caption{Quantitative results of raindrop removal.}
\label{tab:raindrop}
\resizebox{.8\linewidth}{!}{
\begin{tabular}{lcc}
	\hline
	Method & PSNR--$Y_P$ & SSIM--$Y_P$  \\
	\hline
	Eigen13         &         28.59  &         0.6726  \\
	Pix2Pix             &         30.14  &         0.8299  \\
	DeRaindrop (No GAN) &         29.25  &         0.7853  \\
	DeRaindrop          & \textbf{31.57} &         0.9023  \\
    \rowcolor{lightgray} BPP                  &         30.85  & \textbf{0.9180} \\
	\hline
\end{tabular}
}
\end{table}

\begin{figure}
  \centering
  \includegraphics[width=\linewidth]{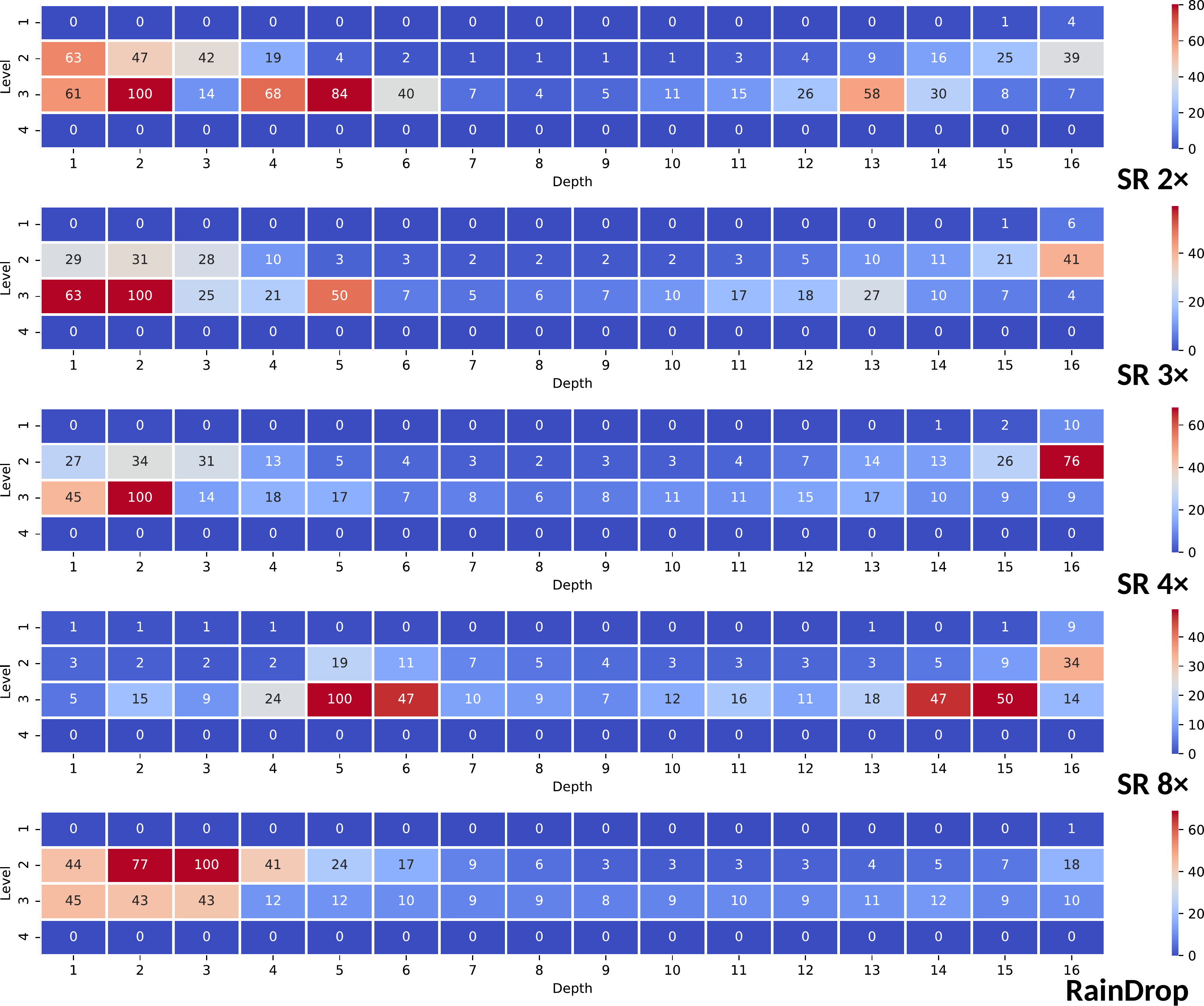}
  \caption{Average $L2$--magnitudes of residual updates normalized by the maximum update with fixed value $100$. \label{fig:residuals_magnitudes}}
\end{figure}

\textbf{Inspection of ODE updates}. We conduct experiments to measure the magnitude of the updates in equation \eqref{eq:bpp_dynamic} to better understand the dynamic of the network when solving different problems. The arrange of Flux units in BPP networks forms an array of size $L\times D$ (number of levels times depth) and we compare the magnitude of residual updates in each one of these units. In Figure \ref{fig:residuals_magnitudes} we display the result of measuring
\begin{equation}
  \left\|\frac{dh_k}{dt}\right\|_2 = \|P_k(R_k(h_k^t, t), h_{k-1}^{t+1}, t)\|_2 \;,
\end{equation}
for every flux unit, averaged over all images in the validation sets, and normalized to the maximum value (fixed to $100$). At the lowest resolution ($k=4$) the reference image never changes and thus the updates is always zero.

Interestingly, we observe that the dynamic is far from the original contraction mapping design of IBP, that would result in an exponential decay of updates along depth. Here, we should remember that the dynamic is driven exclusively by the result of training the network in supervised manner. Instead of an exponential decay, the network consistently shows a bimodal statistic with one peak very close to the input and another very close to the output. Also, the highest resolution receives very small updates meaning that these feature move more or less unchanged with an increased update towards the end. The major processing goes on at the middle levels. In SR updates are stronger at the lower resolution ($k=3$) and for RainDrop removal updates are stronger at the higher resolution ($k=2$). The bimodal statistic is reminiscent of interpretability results for VGG networks in classification, that show higher contribution to label outputs very early and very late in a sequential configuration \cite{LinearScopes}. Nevertheless, in BPP the updates focus on one or two resolution levels as opposed to VGG networks that are designed to process high resolutions early in the network and very low resolutions towards the end. Despite this important difference, these results suggest that sequential networks find solutions in two steps: analysis at the first layers, and fusion towards the very end.
\begin{figure}
  \centering
  \includegraphics[width=\linewidth]{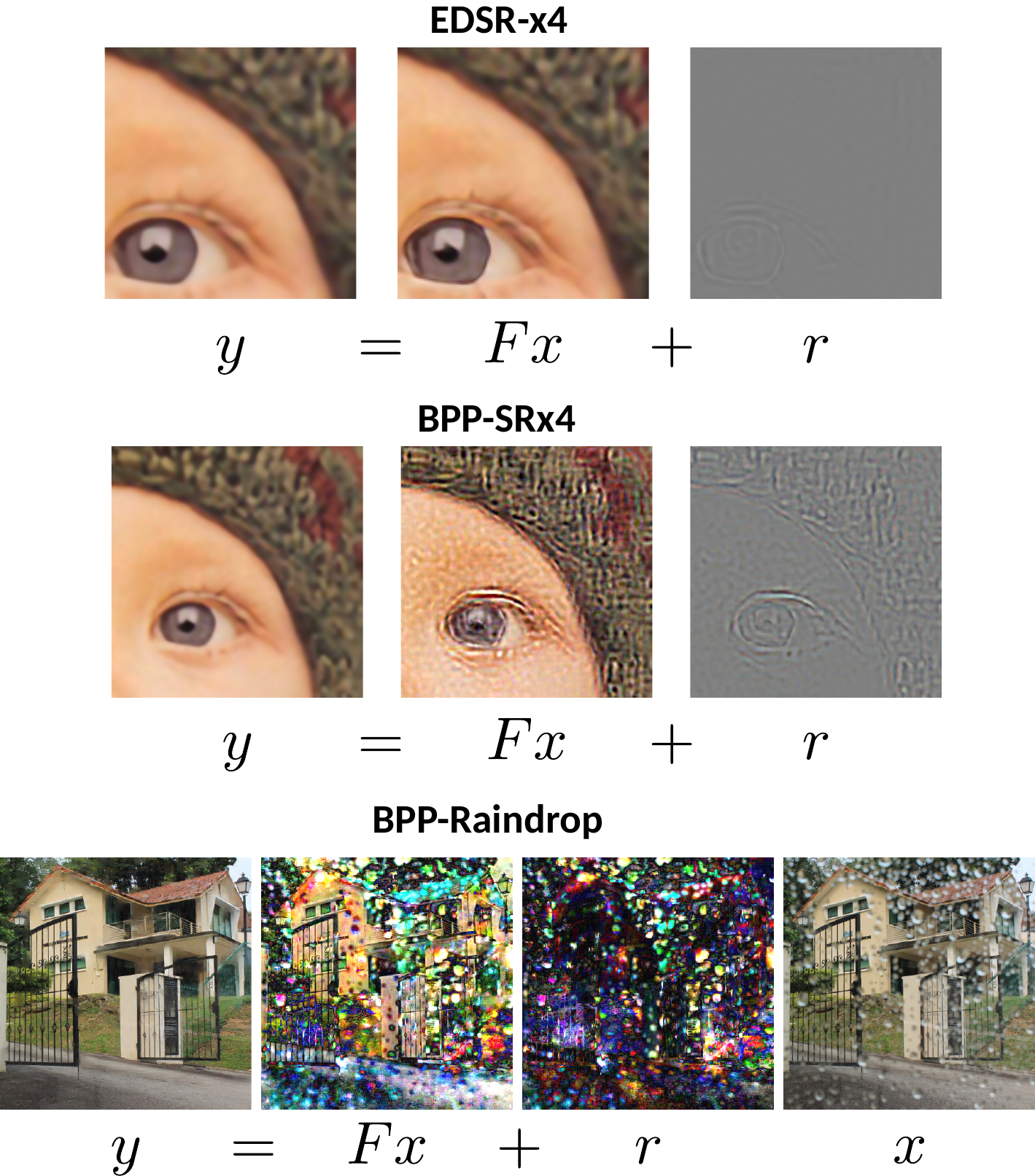}
  \caption{Local and global contributions ($Fx$ and $r$) for $3$ systems using deep filter visualization\cite{PNavarrete_2019a}. EDSR relies on local contributions while BPP balances both local and global contributions.\label{fig:residuals}}
\end{figure}

\textbf{Interpretability}. We apply the \emph{LinearScope} method from \cite{LinearScopes} to analyze the learning process in global and local problems. The general methodology is as follows. The BPP architecture contains several non--linear modules consisting on ReLUs and IN-layers. The decision of which pixels pass or stop in ReLUs, and what mean and variance to use in IN--layers, is non--linear. But the action of these layers are linear: masking and normalizing. For a given input image $x$, the action of all non--linear modules (ReLU and IN--layers) can be fixed as: $1/0$--masks for ReLU and fixed mean and variance in IN--layers. This gives a linear system of the form $y=Fx+r$ that generates the same output as the non--linear system for the input $x$, and represents the overall action of the network on the input image.

The matrix $F$ represents the interpolation filters used by the network to solve the problem, and thus shows the \emph{local processing}. The residual $r$ is a fixed \emph{global} image created by non--linear modules. Figure \ref{fig:residuals} shows the local contributions, $Fx$, and global contributions, $r$, for three systems. We observe that EDSR almost purely relies on local processing to obtain an output. BPP, on the other hand, relies mostly on local processing but the contribution of $r$ is much larger than the one in EDSR. This shows a significantly different approach followed by BPP, compared to EDSR, to solve the super--resolution problem.

The BPP system for raindrop removal reveals a much larger contribution of $r$, that resembles a mask of raindrops. This means that BPP uses a local approach on areas without raindrops (using $Fx$) and a global approach on raindrops determined by the residual $r$. The mechanism used by the network to obtain the residual $r$ is non--linear. Overall, we observe that for this problem the BPP network divides the problem in two parts: a local adaptive filter in clean areas, to nearly copy--paste the input into the output; and a non--linear global approach to fill--in raindrop areas.

\section{Conclusions}
We propose Back--Projection Pipeline as a simple yet non--trivial extension of residual networks (ResNets) to run in multiple resolutions. The update dynamic through the layers of the network includes interactions between different resolutions in a way that is causal in scale, and it is represented by a system of ODEs. We use it as a generic multi--resolution approach to enhance images.
The focus of our investigation is to evaluate this multi--scale residual approach. Overall, our empirical results show that BPP can achieve excellent results in traditional supervised learning. Our BPP configuration gets state--of--the--art results in SR for fixed upscaling factors and competitive results for raindrop removal as well as other problems (see Appendix).
We also observe a lack of generalization for the problem of SR with unknown upscaling factors. Inspection of the residual updates in the network shows that all resolution levels are being used, with higher intensity in lower resolutions, showing that supervised training gives preference to the multi--scale setting over traditional residual networks. Based on our results, we cannot conclude that scale causality is beneficial. Nevertheless, we can at least conclude that this strong simplification in the flow of network information, inherited from IBP, does not prevent the architecture to achieve competitive results. Further investigation is necessary in this regard (especially regarding generalization) and it could open interesting research directions in network architecture search and design.

\begin{figure*}[ht]
    \centering
    \includegraphics[width=\linewidth]{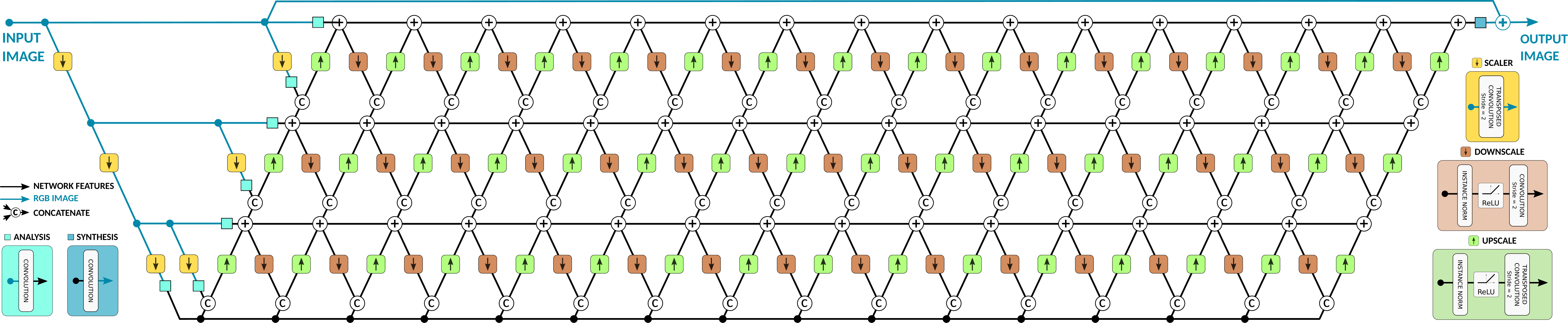}
    \captionof{figure}{Detail diagram of the $4$--level, $16$--layers BPP configuration used in our experiments.}
    \label{fig:bpp_full}
\end{figure*}

{\small
\bibliographystyle{aaai21}
\bibliography{bibliography}
}

\section{Appendix}

\def\thesection{\Alph{section}}
\def\thesubsection{\Alph{section}.\arabic{subsection}}

\vspace{-.1in}
\subsection{Diagrams}
\begin{figure}
  \centering
  \includegraphics[width=.8\linewidth]{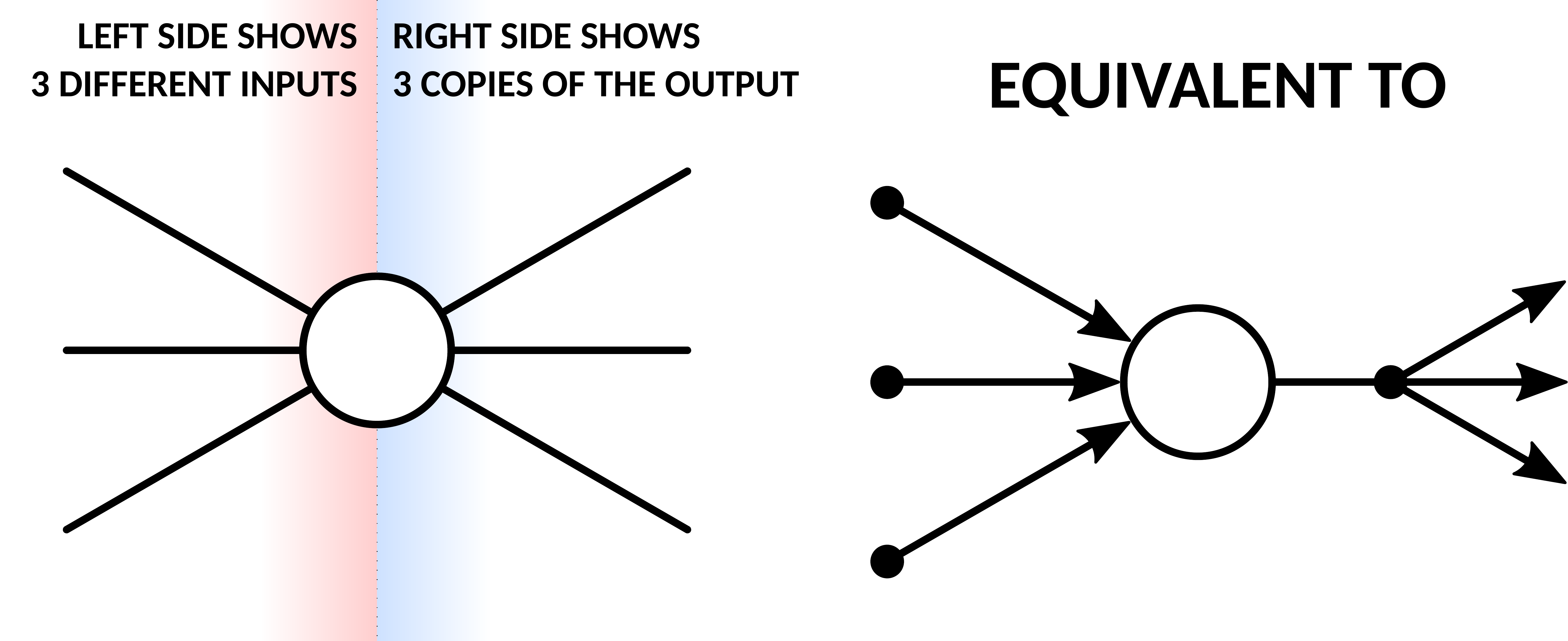}
  \caption{Diagram notation. \label{fig:notation}}
\end{figure}

In an effort to make diagrams easy to read, concise and carrying a precise meaning, we introduce the notation in Figure \ref{fig:notation}. This is, lines connected to the left--side of any given module represent different inputs to that module. Every module can have several inputs but only one output. Lines connected to the right--side of a given module represent copies of the same output.

Figure \ref{fig:bpp_full} shows an expanded diagram of the single BPP configuration used in our experiments. It uses $16$ back--projection layers (flux blocks), $4$ resolution levels and $256$, $128$, $64$ and $48$ features per level from lowest to highest resolution, respectively. All convolutional layers use $3\times 3$ as kernel size, and scalers are initialized with bicubic filters of size $9\times 9$ and trained as additional parameters.

We observe that, after initialization, the lowest--resolution network state (at the bottom of the diagram) never changes. Thus, the highest--resolution state (at the top of the diagram) is always $3$--layers away from this fixed state. This is similar to a long--range skip--connection in DenseNet~\cite{huang2017densely}, but in BPP these shortcut moves through a different resolution. Because of scale causality, the next low--resolution level moves relatively close to the fixed state and we can interpret it as a shorter--range skip--connection. Thus, the particular structure of BPP allows quick paths from the output to every layer of the network, similar to DenseNets, which is convenient for the gradient flow during back--propagation steps.

\subsection{Evaluation Metrics}
Quantitative evaluations in our experiments include three objective metrics: PSNR, SSIM and HIGRADE--2. From these, PSNR and SSIM are reference--based metrics that measure the difference between an impaired image and ground truth. Higher values are better in both cases. The PSNR (range $0$ to $\infty$) is a log--scale version of mean--square--error and SSIM (range $0$ to $1$) uses image statistics to better correlate with human perception. Full expressions are as follows:
\begin{align}
    PSNR(X,Y) & = 10 \cdot \log_{10}\left(\frac{255^2}{MSE}\right) \;,\\
    SSIM(X,Y) & =\frac{(2\mu_X\mu_Y+c_1)(2\sigma_{XY}+c_2)}{(\mu_X^2+\mu_Y^2+c_1)(\sigma_X^2+\sigma_Y^2+c_2)} \;,
\end{align}
where $MSE=\mathbb{E}\left[(X-Y)^2\right]$ is the mean square error of the difference between $X$ and $Y$; $\mu_X$ and $\mu_Y$ are the averages of $X$ and $Y$, respectively; $\sigma_X^2$ and $\sigma_Y^2$ are the variances of $X$ and $Y$, respectively; $\sigma _{XY}$ is the covariance of X and Y; $c_1=6.5025$ and $c_2=58.5225$.

HIGRADE--2\cite{higrade1-2} is a non--reference image quality metric based on gradient scene--statistics defined in the LAB color space and it is often used to evaluate high--dynamic--range images. Here, we used the Matlab code available in \cite{higrade1-2-code}.

In the case of PSNR and SSIM metrics, we follow existing benchmarks that use different versions of these metrics. We used the following three definitions in our experiments:
\begin{itemize}
    \item $\boldsymbol{PSNR/SSIM-Y_M}$: Based on the \textbf{M}atlab code available in \cite{psnr-ssim-y}, computes PSNR/SSIM on the $Y$ channel. Matlab uses a conversion of RGB to YUV color--spaces following the BT.709 standard, including offsets that are often avoided in other implementations.
    \item $\boldsymbol{PSNR/SSIM-Y_P}$: Based on the \textbf{P}ython code available in \cite{deraindrop-evaluation-code}, computes PSNR/SSIM on the $Y$ channel. The code uses an OpenCV function to convert from RGB to YCbCr color--space.
    \item $\boldsymbol{PSNR/SSIM-RGB}$: Based on the Python code available in \cite{ntire19score}, computes the average PSNR/SSIM for pairs of RGB images.
\end{itemize}

\subsection{Training Settings}

\subsubsection{Image Super--Resolution}

We use DIV2K\cite{Agustsson_2017_CVPR_Workshops} and FLICKR--2K datasets for training and the following datasets for test: Set--14\cite{zeyde2010on}, BSDS--100\cite{martin2001a}, Urban--100\cite{huang2015single} and Manga--109\cite{matsui2017sketch-based}. Impaired images were obtained by downscaling and then upscaling ground truth images, using Bicubic scaler, with scaling factors: $2\times$, $3\times$, $4\times$ and $8\times$. Our target is to recover the ground truth so we use a loss function that measures the $L_1$ distance between impaired images and ground truth. For evaluation we measure PSNR and SSIM on the Y--channel using the Matlab code from \cite{psnr-ssim-y}.

We follow the training settings from \cite{Lim_2017_CVPR_Workshops}. In each training batch, we randomly take $16$ impaired patches from our training set ($800$ DIV2K plus $2,650$ FLICKR--2K images). We consider two cases: we train a model BPP--SR$\times f$ for each upscaling factor $f=2, 3, 4$ and $8$; and we also train a model BPP--SRx2x3x4x8 to restore impaired images with unknown upscaling factor. We use patch size $48f\times 48f$, for $f=2, 3$ and $4$, and $192\times 192$ for $f=8$ and unknown upscaling factor. We augment the patches by random horizontal/vertical flipping and rotating $90^\circ$. We use Adam optimizer\cite{kingma2014adam} with learning rate initialized to $10^{-4}$ and decreased by half every $200,000$ back--propagation steps.

The training data used for the BPP--SRx2x3x4x8 model includes all images used for training the upscaling factors $f=2, 3, 4$ and $8$. We could have chosen to train our model using a random and fractional upscaling factor $2.0\leqslant f\leqslant 8.0$, but this would have made it difficult to reproduce the training settings.

\begin{table*}
\caption{Extended quantitative evaluation for super--resolution.} \label{tab:sr_extended}
\centering
\resizebox{\linewidth}{!}{
\begin{tabular}{lccccccccc}
    \hline
    & &
    \multicolumn{2}{c}{Set14} &
    \multicolumn{2}{c}{BSDS100} &
    \multicolumn{2}{c}{Urban100} &
    \multicolumn{2}{c}{Manga109} \\
    Algorithm & & PSNR--$Y_M$ & SSIM--$Y_M$ & PSNR--$Y_M$ & SSIM--$Y_M$ & PSNR--$Y_M$ & SSIM--$Y_M$ & PSNR--$Y_M$ & SSIM--$Y_M$ \\
    \hline
    Bicubic & \multirow{16}{*}{$2\times$}                          &  30.34  &  0.870  &  29.56  &  0.844  &  26.88  &  0.841  &  30.84  &  0.935 \\
    A+~\cite{Timofte_2014a}                                      & &  32.40  &  0.906  &  31.22  &  0.887  &  29.23  &  0.894  &  35.33  &  0.967 \\
    FSRCNN~\cite{Dong_2016a}                                     & &  32.73  &  0.909  &  31.51  &  0.891  &  29.87  &  0.901  &  36.62  &  0.971 \\
    SRCNN~\cite{Dong_2014a}                                      & &  32.29  &  0.903  &  31.36  &  0.888  &  29.52  &  0.895  &  35.72  &  0.968 \\
    MSLapSRN~\cite{MSLapSRN}                                     & &  33.28  &  0.915  &  32.05  &  0.898  &  31.15  &  0.919  &  37.78  &  0.976 \\
    VDSR~\cite{Kim_2016_VDSR}                                    & &  32.97  &  0.913  &  31.90  &  0.896  &  30.77  &  0.914  &  37.16  &  0.974 \\
    LapSRN~\cite{LapSRN}                                         & &  33.08  &  0.913  &  31.80  &  0.895  &  30.41  &  0.910  &  37.27  &  0.974 \\
    DRCN~\cite{Kim_2016_DRCN}                                    & &  32.98  &  0.913  &  31.85  &  0.894  &  30.76  &  0.913  &  37.57  &  0.973 \\
    MGBP~\cite{PNavarrete_2019a}                                 & &  33.27  &  0.915  &  31.99  &  0.897  &  31.37  &  0.920  &  37.92  &  0.976 \\
    D-DBPN~\cite{DBPN2018}                                       & &  33.85  &  0.919  &  32.27  &  0.900  &  32.70  &  0.931  &  39.10  &  0.978 \\
    EDSR~\cite{Lim_2017_CVPR_Workshops}                          & &  33.92  &  0.919  &  32.32  &  0.901  &  32.93  &  0.935  &  39.10  &  0.977 \\
    RDN~\cite{zhang2018rdnir}                                    & & \textbf{34.28} & \textbf{0.924} & \textbf{32.46} & \textbf{0.903} & \textbf{33.36} & \textbf{0.939} & \textbf{39.74} & \textbf{0.979} \\
    RCAN~\cite{zhang2018rcan}                                    & &  34.12  &  0.921  &  32.41  & \textbf{0.903} &  33.34  &  0.938  &  39.44  & \textbf{0.979} \\
    \rowcolor{lightgray} BPP--SRx2x3x4x8                         & &  33.27  &  0.913  &  31.21  &  0.879  &  31.67  &  0.921  &  38.31  &  0.975  \\
    \rowcolor{lightgray} BPP--SRx2                               & &  34.23  &  0.922  &  31.63  &  0.886  &  33.07  &  0.935  &  39.19  &  0.977  \\
    \noalign{\smallskip}
    \hline
    \noalign{\smallskip}
    Bicubic & \multirow{8}{*}{$3\times$}                           &  27.55  &  0.774  &  27.21  &  0.739  &  24.46  &  0.735  &  26.95  &  0.856 \\
    SRCNN~\cite{Dong_2014a}                                      & &  29.30  &  0.822  &  28.41  &  0.786  &  26.24  &  0.799  &  30.48  &  0.912 \\
    MSLapSRN~\cite{MSLapSRN}                                     & &  29.97  &  0.836  &  28.93  &  0.800  &  27.47  &  0.837  &  32.68  &  0.939 \\
    LapSRN~\cite{LapSRN}                                         & &  29.87  &  0.832  &  28.82  &  0.798  &  27.07  &  0.828  &  32.21  &  0.935 \\
    EDSR~\cite{Lim_2017_CVPR_Workshops}                          & &  30.52  &  0.846  &  29.25  &  0.809  &  28.80  &  0.865  &  34.17  &  0.948 \\
    RDN~\cite{zhang2018rdnir}                                    & &  30.74  & \textbf{0.850} & \textbf{29.38} & \textbf{0.812} &  29.18  &  0.872  & \textbf{34.81} & \textbf{0.951} \\
    \rowcolor{lightgray} BPP--SRx2x3x4x8                         & &  30.23  &  0.838  &  28.81  &  0.794  &  28.43  &  0.852  &  33.75  &  0.943  \\
    \rowcolor{lightgray} BPP--SRx3                               & &  \textbf{30.78}  &  0.848  &  29.14  &  0.804  &  \textbf{29.56}  &  \textbf{0.873}  &  34.49  &  0.948  \\
    \noalign{\smallskip}
    \hline
    \noalign{\smallskip}
    Bicubic & \multirow{16}{*}{$4\times$}                          &  26.10  &  0.704  &  25.96  &  0.669  &  23.15  &  0.659  &  24.92  &  0.789 \\
    A+~\cite{Timofte_2014a}                                      & &  27.43  &  0.752  &  26.82  &  0.710  &  24.34  &  0.720  &  27.02  &  0.850 \\
    FSRCNN~\cite{Dong_2016a}                                     & &  27.70  &  0.756  &  26.97  &  0.714  &  24.61  &  0.727  &  27.89  &  0.859 \\
    SRCNN~\cite{Dong_2014a}                                      & &  27.61  &  0.754  &  26.91  &  0.712  &  24.53  &  0.724  &  27.66  &  0.858 \\
    MSLapSRN~\cite{MSLapSRN}                                     & &  28.26  &  0.774  &  27.43  &  0.731  &  25.51  &  0.768  &  29.54  &  0.897 \\
    VDSR~\cite{Kim_2016_VDSR}                                    & &  28.03  &  0.770  &  27.29  &  0.726  &  25.18  &  0.753  &  28.82  &  0.886 \\
    LapSRN~\cite{LapSRN}                                         & &  28.19  &  0.772  &  27.32  &  0.728  &  25.21  &  0.756  &  29.09  &  0.890 \\
    DRCN~\cite{Kim_2016_DRCN}                                    & &  28.04  &  0.770  &  27.24  &  0.724  &  25.14  &  0.752  &  28.97  &  0.886 \\
    MGBP~\cite{PNavarrete_2019a}                                 & &  28.43  &  0.778  &  27.42  &  0.732  &  25.70  &  0.774  &  30.07  &  0.904 \\
    D-DBPN~\cite{DBPN2018}                                       & &  28.82  &  0.786  &  27.72  &  0.740  &  26.54  &  0.795  &  31.18  &  0.914 \\
    EDSR~\cite{Lim_2017_CVPR_Workshops}                          & &  28.80  &  0.788  &  27.71  &  0.742  &  26.64  &  0.803  &  31.02  &  0.915 \\
    RDN~\cite{zhang2018rdnir}                                    & &  29.01  & \textbf{0.791} &  27.85  & \textbf{0.745} &  27.01  &  0.812  & \textbf{31.74} & \textbf{0.921} \\
    RCAN~\cite{zhang2018rcan}                                    & &  28.87  &  0.789  &  27.77  &  0.744  &  26.82  &  0.809  &  31.22  &  0.917 \\
    \rowcolor{lightgray} BPP--SRx2x3x4x8                         & &  28.55  &  0.778  &  27.43  &  0.728  &  26.48  &  0.791  &  30.81  &  0.909  \\
    \rowcolor{lightgray} BPP--SRx4                               & &  \textbf{29.07}  &  \textbf{0.791}  &  \textbf{27.89}  &  \textbf{0.745}  &  \textbf{27.55}  &  \textbf{0.819}  &  31.63  &  0.918  \\
    \noalign{\smallskip}
    \hline
    \noalign{\smallskip}
    Bicubic & \multirow{15}{*}{$8\times$}                          &  23.19  &  0.568  &  23.67  &  0.547  &  20.74  &  0.516  &  21.47  &  0.647 \\
    A+~\cite{Timofte_2014a}                                      & &  23.98  &  0.597  &  24.20  &  0.568  &  21.37  &  0.545  &  22.39  &  0.680 \\
    FSRCNN~\cite{Dong_2016a}                                     & &  23.93  &  0.592  &  24.21  &  0.567  &  21.32  &  0.537  &  22.39  &  0.672 \\
    SRCNN~\cite{Dong_2014a}                                      & &  23.85  &  0.593  &  24.13  &  0.565  &  21.29  &  0.543  &  22.37  &  0.682 \\
    MSLapSRN~\cite{MSLapSRN}                                     & &  24.57  &  0.629  &  24.65  &  0.592  &  22.06  &  0.598  &  23.90  &  0.759 \\
    VDSR~\cite{Kim_2016_VDSR}                                    & &  24.21  &  0.609  &  24.37  &  0.576  &  21.54  &  0.560  &  22.83  &  0.707 \\
    LapSRN~\cite{LapSRN}                                         & &  24.44  &  0.623  &  24.54  &  0.586  &  21.81  &  0.582  &  23.39  &  0.735 \\
    MGBP~\cite{PNavarrete_2019a}                                 & &  24.82  &  0.635  &  24.67  &  0.592  &  22.21  &  0.603  &  24.12  &  0.765 \\
    D-DBPN~\cite{DBPN2018}                                       & &  25.13  &  0.648  &  24.88  &  0.601  &  22.83  &  0.622  &  25.30  &  0.799 \\
    EDSR~\cite{Lim_2017_CVPR_Workshops}                          & &  24.94  &  0.640  &  24.80  &  0.596  &  22.47  &  0.620  &  24.58  &  0.778 \\
    RDN~\cite{zhang2018rdnir}                                    & &  25.38  &  0.654  &  25.01  &  0.606  &  23.04  &  0.644  &  \textbf{25.48}  &  \textbf{0.806} \\
    RCAN~\cite{zhang2018rcan}                                    & &  25.23  &  0.651  &  24.98  &  0.606  &  23.00  &  0.645  &  25.24  &  0.803 \\
    \rowcolor{lightgray} BPP--SRx2x3x4x8                         & &  25.10  &  0.642  &  24.89  &  0.598  &  22.72  &  0.626  &  24.78  &  0.785  \\
    \rowcolor{lightgray} BPP--SRx8                               & &  \textbf{25.53}  &  \textbf{0.655}  &  \textbf{25.11}  &  \textbf{0.607}  &  \textbf{23.17}  &  \textbf{0.649}  &  25.28  &  0.800  \\
    \hline
    \end{tabular}
}
\end{table*}

\begin{figure*}
  \centering
  \includegraphics[width=\linewidth]{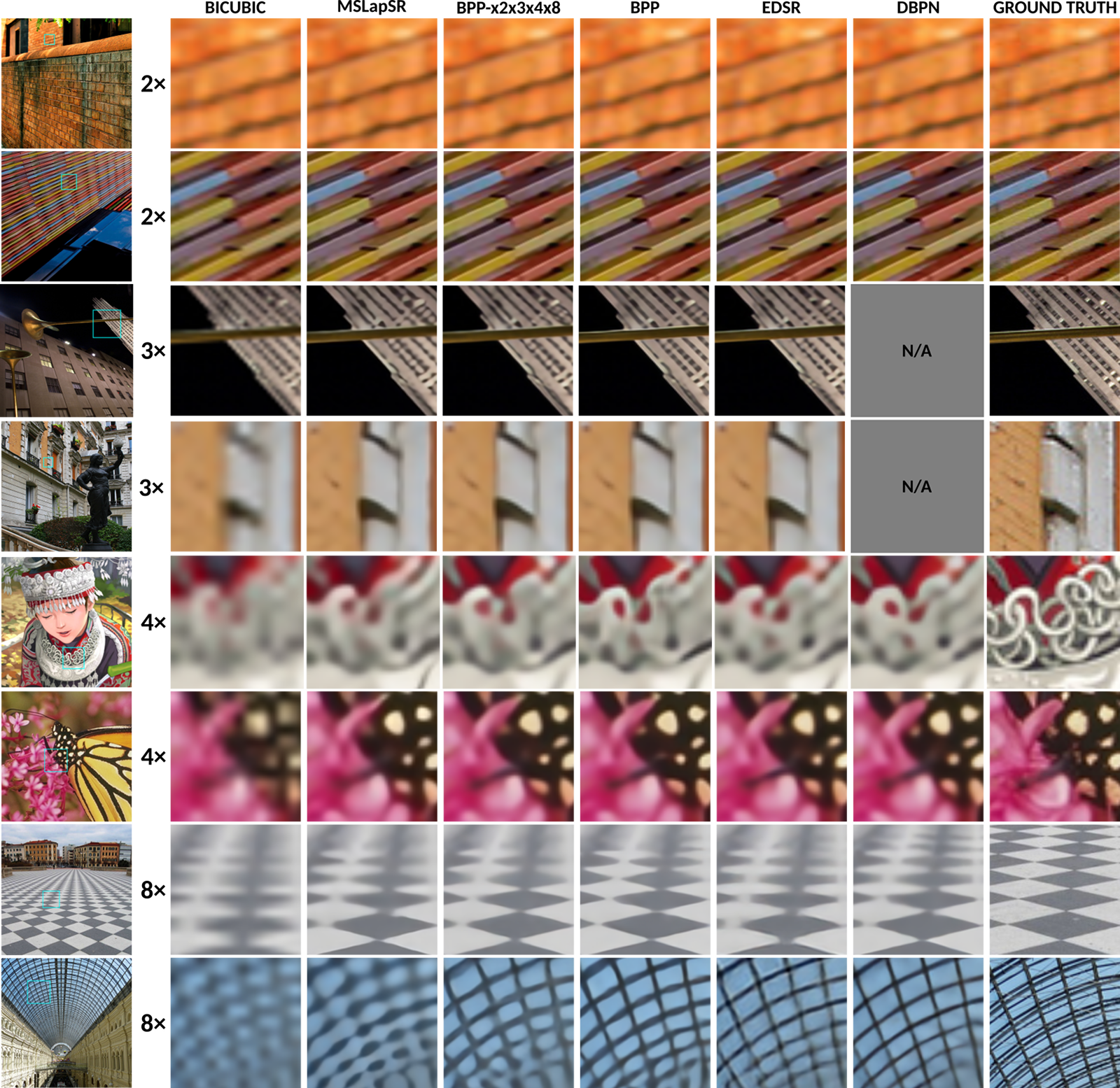}
  \caption{Extended qualitative evaluation for super--resolution. \label{fig:sr_extended}}
\end{figure*}

\subsubsection{Mobile--to--DSLR Photo Translation}
 We use the DPED\cite{ignatov2017dslr} dataset for training and test. This dataset provides $100\times 100$ aligned patches taken from iPhone--mobile photos (impaired) and DSLR--Canon photos (ground truth). There are $160,471$ patches available for training and $4,353$ patches for test. We take $400$ patches from the test set for validation during training. We use full size iPhone images from DPED for qualitative results. For loss function we use the negative SSIM between impaired and ground truth patches. We find SSIM to be more effective than $L_1$ and MSE losses in this problem. For evaluation we measure the average PSNR and SSIM metrics for RGB pairs, using the code from \cite{ntire19score}, and the non--reference metric HIGRADE--2\cite{higrade1-2} using the Matlab code available from \cite{higrade1-2-code}.

In each training batch, we take $16$ patches of size $100\times 100$. We use Adam optimizer\cite{kingma2014adam} with learning rate initialized to $10^{-4}$ and decreased by half every $200,000$ back--propagation steps. We do not observe improvements after $200$ epochs.

\begin{figure*}
  \centering
  \includegraphics[width=\linewidth]{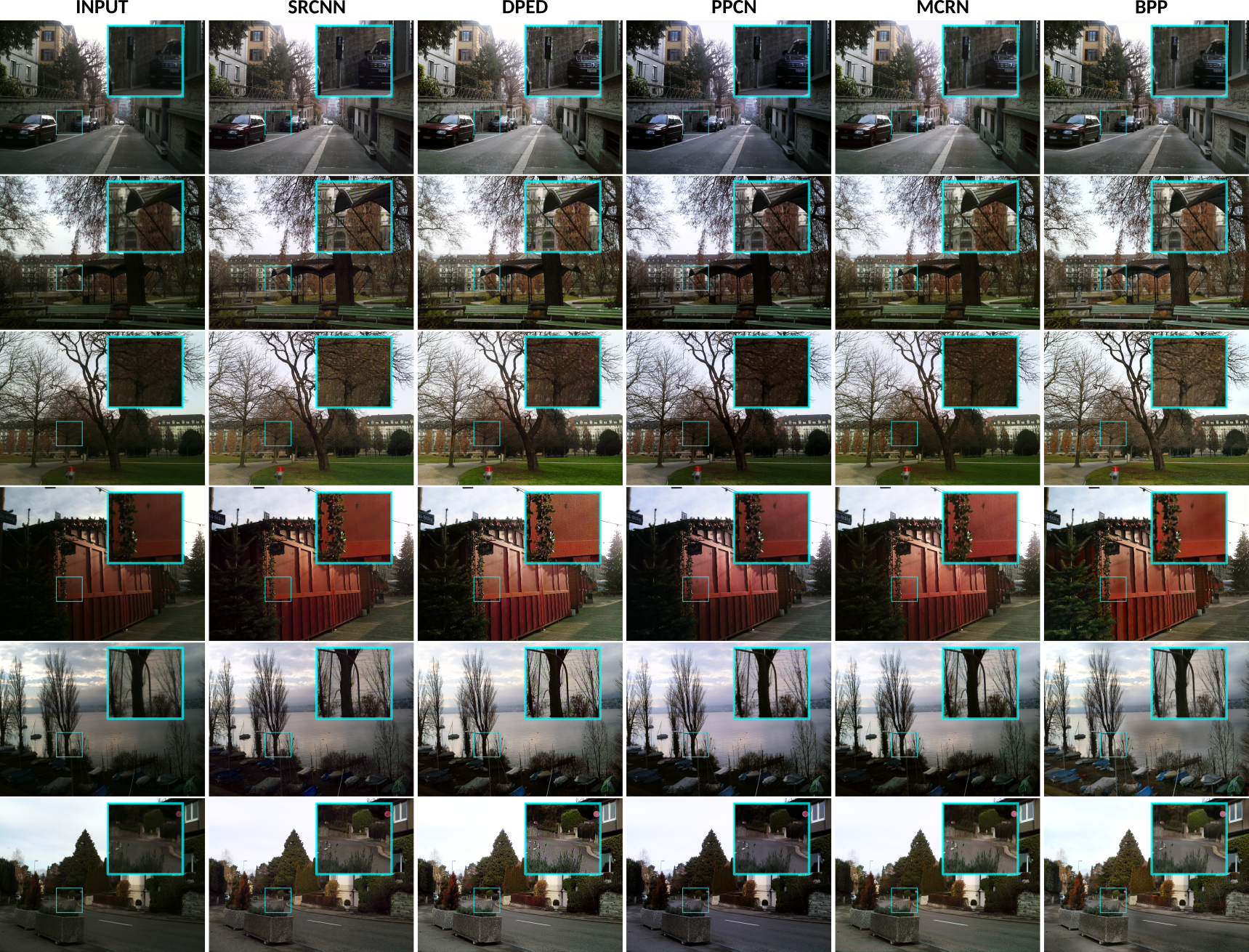}
  \caption{Extended qualitative evaluation for Mobile--to--DSLR photo translation. \label{fig:dped_extended}}
\end{figure*}

\subsubsection{Image Dehaze}

We use the following real haze datasets: I--Haze\cite{ancuti2018ihaze}, O--Haze\cite{ancuti2018ohaze} and Dense--Haze\cite{DenseHaze}. We follow the training setting from \cite{dehaze_zhang_2018w}. In each training batch, we take $1$ patch of size $528\times 528$. The training set is augmented by rescaling the images, using bicubic scaler, to $1.25\times$, $1\times$, $0.625\times$ and $0.3125\times$ the original size. We use Adam optimizer\cite{kingma2014adam} with learning rate initialized to $10^{-4}$ and decreased by half every $200,000$ back--propagation steps. We train the system for $10,000$ epochs.

\begin{figure*}
  \centering
  \includegraphics[width=\linewidth]{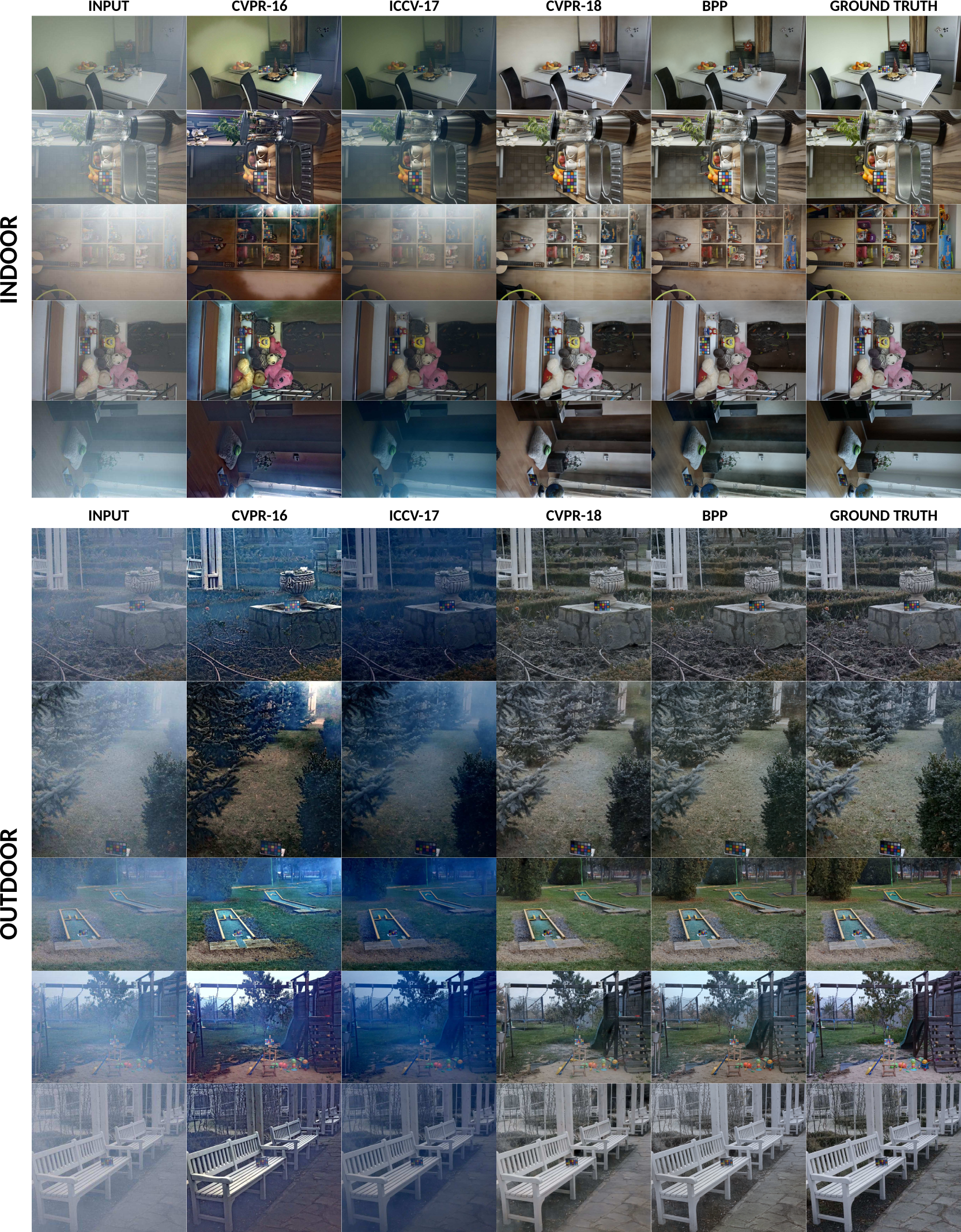}
  \caption{Extended qualitative evaluation for image dehaze for Indoor/Outdoor datasets. \label{fig:dehaze_extended_a}}
\end{figure*}
\begin{figure*}
  \centering
  \includegraphics[width=\linewidth]{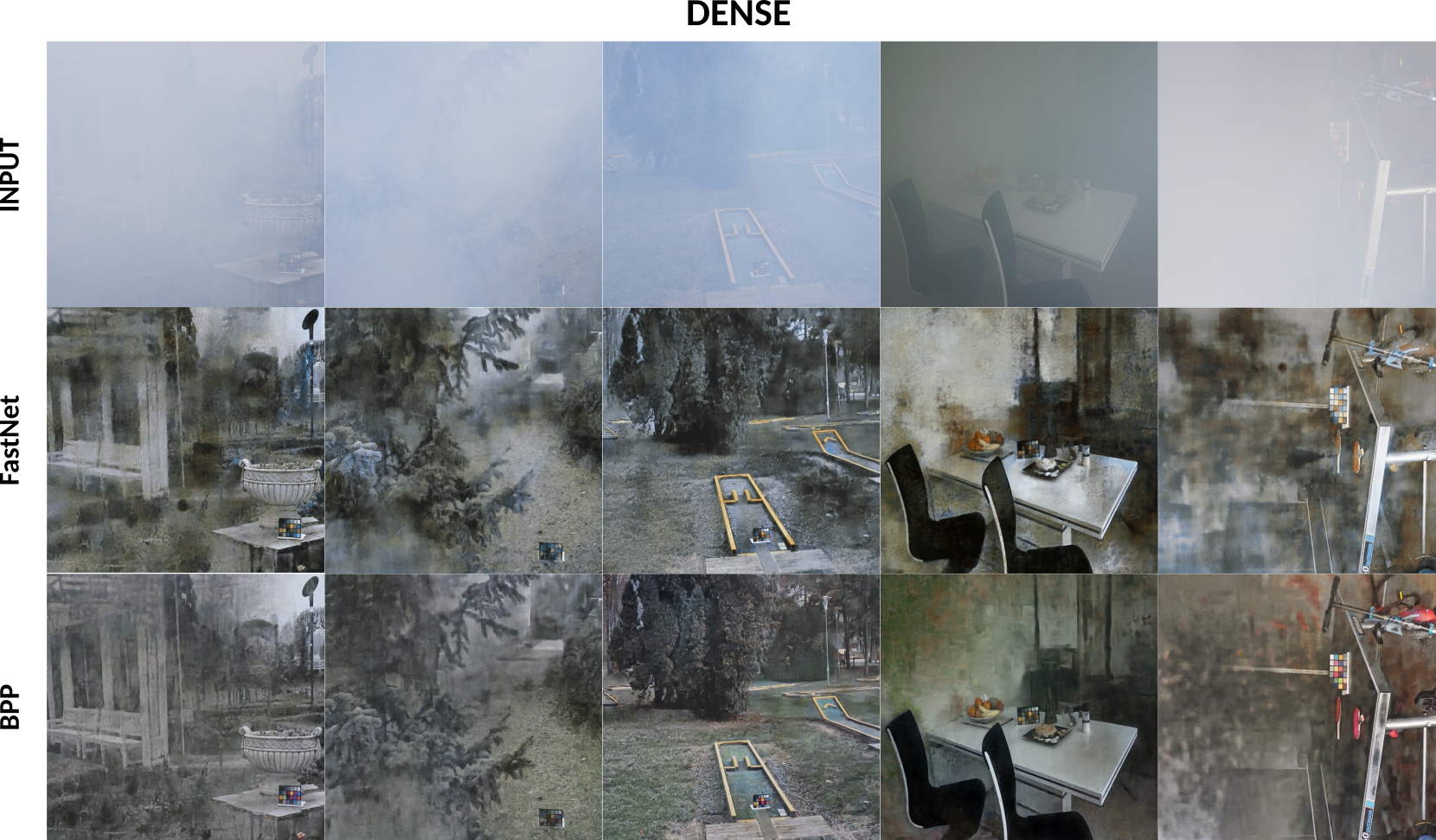}
  \caption{Extended qualitative evaluation for image dehaze for Dense dataset. \label{fig:dehaze_extended_b}}
\end{figure*}

\subsubsection{Joint HDR and Super--Resolution}

We use the HDR--Eye\cite{nemoto2015visual} dataset for training and Wang LDR\cite{wang2013naturalness} dataset for test. HDR--Eye\cite{nemoto2015visual} provides HDR images constructed from multi--exposure photographs. Following the training configuration in \cite{soh_joint_2019}, we select $40$ from a total of $46$ standard--exposed and HDR--constructed pairs of images (we excluded images C01.png, C04.png, C13.png, C28.png, C38.png and C42.png, because of visible misalignment problems in the HDR image constructions). Then, we take each standard--exposed image and we: first, downscale it by factor $2$; and then upscale it by factor $2$ (both with bicubic scaler), and use this output as impaired image. We follow the configuration in \cite{soh_joint_2019} although their tone--mapping algorithms are not specified and tone--mapped images are not provided. We use several tone--mapping algorithms until being able to produce competitive quantitative and qualitative outputs. For our final results we used the OpenCV implementation of Reinhard--Devlin tone--mapping~\cite{reinhard2005dynamic} with parameters $\text{gamma}=2.2$, $\text{intensity}=0$, $\text{light\_adapt}=0.$ and $\text{color\_adapt}=0$.

We train our system using patches of size $456\times 456$. Following the analysis in \cite{soh_joint_2019}, we use the non--reference image quality metrics: Ma\cite{Ma-Metric-2017,ma-evaluation-code}, to evaluate SR improvements; and HIGRADE--2\cite{higrade1-2,higrade1-2-code} to evaluate HDR improvements. We augment the patches by random horizontal/vertical flipping and rotating $90^\circ$. We use Adam optimizer\cite{kingma2014adam} with learning rate initialized to $10^{-4}$ and decreased by half every $200,000$ back--propagation steps. We train the system for $10,000$ epochs.

\begin{figure*}
  \centering
  \includegraphics[width=\linewidth]{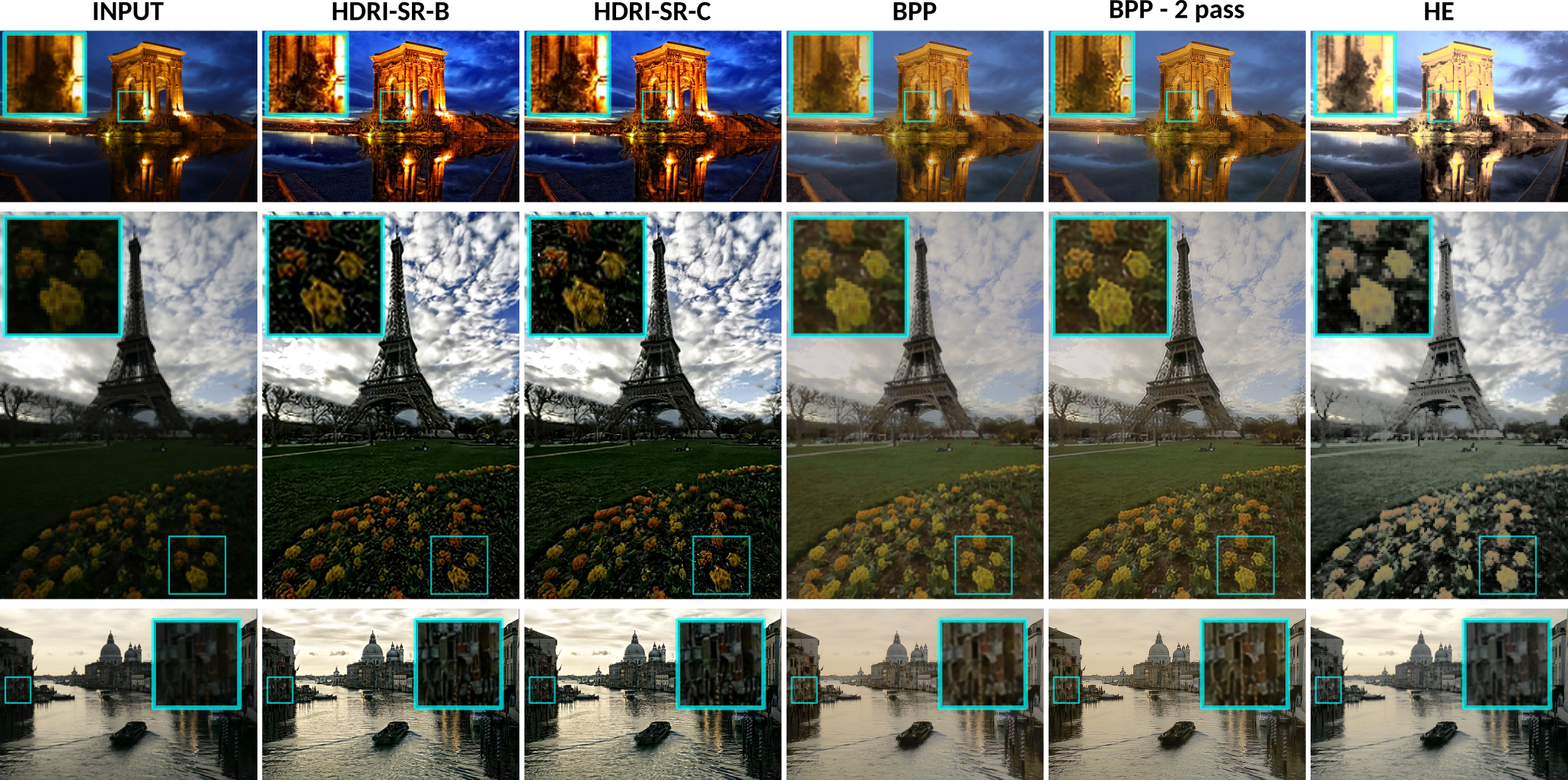}
  \caption{Extended qualitative evaluation for joint HDR+SR enhancement. \label{fig:hdrsr_extended}}
\end{figure*}

\subsubsection{Raindrop Removal}

 We use the DeRaindrop\cite{Qian_2018_CVPR,deraindrop-dataset} dataset for training and test. This dataset provides paired images, one degraded by raindrops and the other one free from raindrops. In each training batch, we take $1$ patch of size $528\times 528$. We train a BPP model using $L_1$ loss and patch size $456\times 456$. We use Adam optimizer\cite{kingma2014adam} with learning rate initialized to $10^{-4}$ and decreased by half every $200,000$ back--propagation steps. We train the system for $10,000$ epochs.

\begin{figure*}
  \centering
  \includegraphics[width=\linewidth]{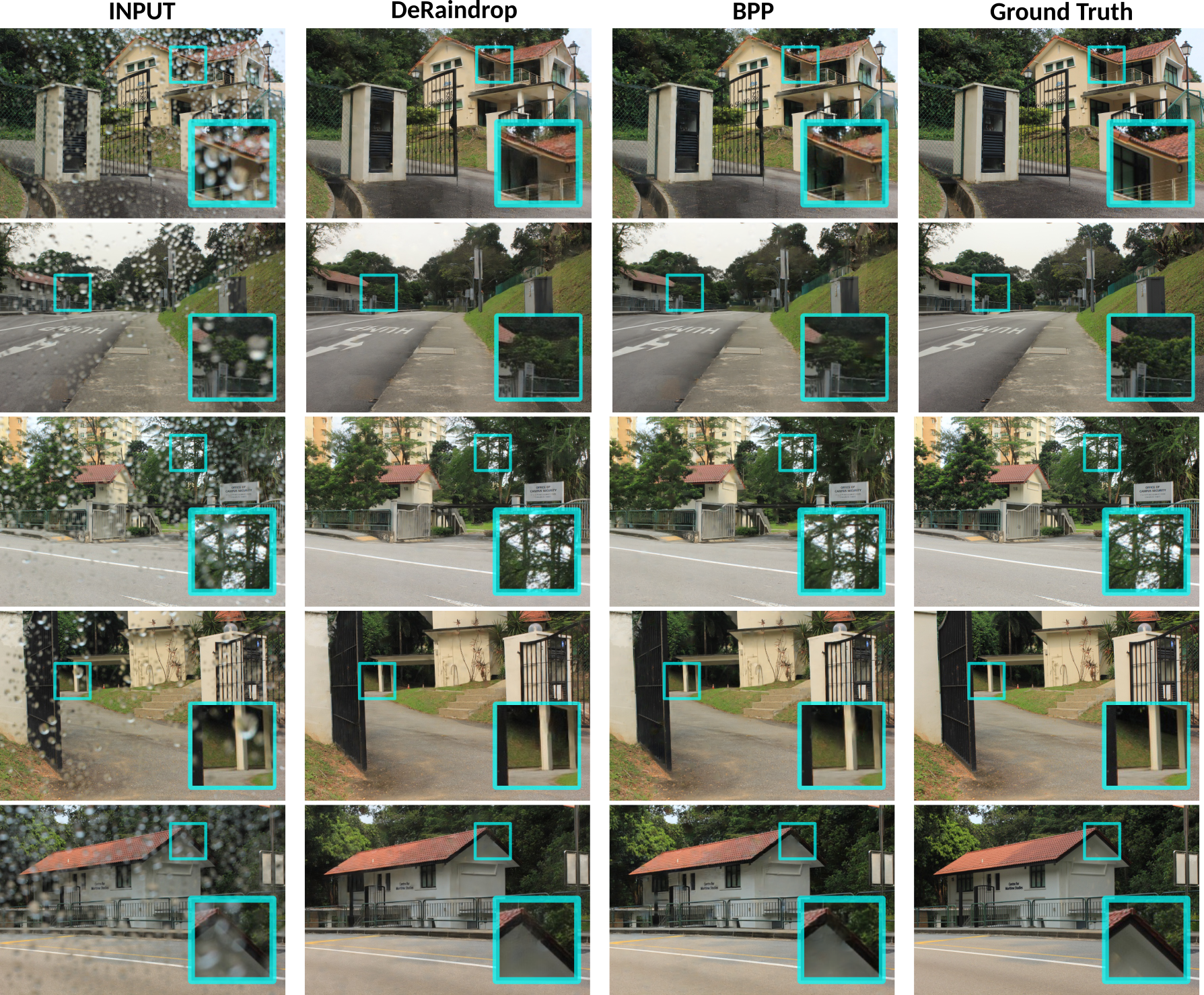}
  \caption{Extended qualitative evaluation for raindrop removal. \label{fig:raindrop_extended}}
\end{figure*}

\subsection{Computing Infrastructure}
All training processes run on Linux operating system, using implementations in Python language with software packages: Numpy, Pytorch\cite{paszke2017automatic}, Scilab, Pillow and OpenCV. We used NVIDIA Tesla M40 (24GB) GPU for training and NVIDIA Titan--X Maxwell (12GB) for tests.

\subsection{Reproducibility}
All output images of the BPP systems obtained in our experiments can be downloaded from the following
\href{https://www.dropbox.com/s/3w0j1bxd2mjkokc/output_images.zip}{link (2.77 GB)}
This can be used to reproduce all quantitative evaluations in our experiments. We have provided external links to all evaluation scripts used in our evaluations.

\end{document}